%% file: ms.tex
\documentclass{emulateapj}
\usepackage{epsfig}
\usepackage{epsf}
\newcommand{\beq}{\begin{equation}}
\newcommand{\eeq}{\end{equation}}
\begin{document}
\shorttitle{The Bound Mass of Substructures}
\shortauthors{Shaw et al.}
\title{The Bound Mass of Substructures in Dark Matter Halos}

\author{Laurie D. Shaw\altaffilmark{1,4}}
\affil{Department of Physics, McGill University, Montreal QC H3A
2T8}
\email{lds@ast.cam.ac.uk}
\author{Jochen Weller\altaffilmark{2}}
\affil{Department of Physics and Astronomy, University College
  London, Gower Street, London WC1E 6BT, UK.}
\author{Jeremiah P Ostriker\altaffilmark{1,3} and Paul Bode\altaffilmark{3}}
\affil{Princeton University Observatory,
  Princeton NJ 08544-1001}

\altaffiltext{1}{Institute of Astronomy, University of Cambridge,
  Madingley Road, Cambridge CB30HA} 
\altaffiltext{2}{Department of Physics and Astronomy, University College
  London, Gower Street, London WC1E 6BT, UK.}
\altaffiltext{3}{Princeton University Observatory,
  Princeton NJ 08544-1001} 
\altaffiltext{4}{Department of Physics, McGill University, Montreal QC H3A
2T8}

\begin{abstract}
We present a new definition of subhalos in dissipationless dark matter
N-body simulations, based on the coherent identification of their
dynamically bound constituents. Whereas previous methods of
determining the energetically bound components of a subhalo ignored
the contribution of all the remaining particles in the halo (those not
geometrically or dynamically associated with the subhalo), our method
allows for all the forces, both internal and external, exerted on the
subhalo. We demonstrate, using the output of a simulation at different
timesteps, that our new method is more accurate at identifying the
bound mass of a subhalo. We then compare our new method to previously
adopted means of identifying subhalos by applying each to a sample of
1838 virialized halos extracted from a high resolution cosmological
simulation. We find that the subhalo distributions are similar in each
case, and that the increase in the binding energy of a subhalo from
including all the particles located within it is almost entirely
balanced by the losses due to the external forces; the net increase in
the mass fraction of subhalos is roughly 10\%, and the extra
substructures tending to reside in the inner parts of the
system. Finally, we compare the subhalo populations of halos to the
sub-subhalo populations of subhalos, finding the two distributions to
be similar. This is a new and interesting result, suggesting a
self-similarity in the hierarchy substructures within cluster mass
halos.
\end{abstract}

\section{INTRODUCTION}

In the current paradigm of hierarchical structure formation, satellite
galaxies in clusters are associated with the remnants of dark matter
halos -- known as subhalos -- that have, at some point in their
history, been accreted and absorbed by a more massive halo. Once
captured by their host (or mother) halo, subhalos are continuously
eroded by the combined effects of dynamical friction and tidal
stripping by the host halo core. Often a subhalo will lose a large
fraction of its mass at the time of accretion, with only the dense core
surviving until the present day.  
In studies of structure formation, it is these self-bound remnants of
the original subhalo which we associate with satellite galaxies.

With the advent of high resolution cosmological simulations, we have
been able to test models of hierarchical structure formation in a
$\Lambda$CDM universe. Until the end of the last decade, it was not
possible to reach the required mass resolution to enable the
identification of galaxy mass halos as substructures in clusters
\citep{White:76, vanKampen:95, Summers:95, Moore:96}.  Commonly known
as the {\it overmerging} problem, this was mainly due to the limited
mass and force resolution of the simulations used.  The major causes
of this problem were premature tidal disruption due to inadequate
force resolution and two-particle evaporation for halos with a small
number of particles \citep{Klypin:99}. Over the last 7 years, rapid
advances in parallel computing, through both the improvement of
hardware and the development of fast and efficient parallel
algorithms, has enabled us to achieve the numerical resolution
required to overcome these numerical problems and probe the subhalo
populations of $\Lambda$CDM halos \citep{Ghigna:98, Klypin:99, Moore:99a,
Okamoto:99, Ghigna:00, Bode:01, Springel:01a, Kravtsov:04, DeLucia:04,
Gao:04, Reed:05}.

However, the exact definition of a `subhalo' is somewhat
ambiguous, and it tends to be intrinsically linked to the
algorithm adopted to identify the groups of particles belonging to
each subhalo in a numerical simulation. Therefore, what constitutes a
substructure is often dependent on the process or quantity
being investigated.  For example, if one is interested in using
numerical simulations to investigate the impact of substructures
within gravitational lenses, and therefore sampling the projected
potential field within objects, then a geometrical definition of a
subhalo is sufficient (see, for example, \citet{Hennawi:05, Hagan:05,
Mao:04, Amara:06}). However, if one instead wants to study the
formation and evolution of subhalos as they initially grow through
accretion, are captured by larger structures, and are subsequently
depleted through dynamical friction and tidal ablation, one must then
search for structures in 6 dimensional phase space in order to
identify statistically significant groups of particles
\citep{Taffoni:03, Hayashi:03, Kazantzidis:04, Kravtsov:04, Gill:04b}.

However, there is an additional and often dominant goal in defining
`substructure'. In most assemblages that we call `galaxies', the
lifetimes of the stars that we observe (especially in the K-band) are
quite long compared to their dynamical (or orbital) times in the
systems in which they are found. Hence, if the purpose is to identify
objects that correspond to the galaxies that we observe, then we must
identify those subhalo particles that will remain together over many
dynamical timescales. In this case, all the forces on a subhalo,
internal and external, must be accounted for in order to identify such
objects. Requiring that all particles be energetically bound, as is
adopted by many authors (e.g. \citet{Ghigna:00, Springel:01a,
Kravtsov:04, DeLucia:04, Gao:04, Reed:05}), can be a necessary, but is
not a sufficient, condition for coherence. We will later spell out
what we believe to be a more appropriate algorithm.

There are many existing algorithms for defining and identifying halos
and subhalos in dissipationless N-body simulations. As noted above,
some methods are essentially geometrical, others aim at finding
dynamically coherent structures; all contain both dimensionless and
dimensional parameters.  In general, there are three levels of
refinement that one can adopt. The most basic is to use a geometrical
routine, such as the Friends-Of-Friends \citep{Huchra:82, Davis:85,
Lacey:94} or Denmax \citep{Bertschinger:91, Gelb:94, Eisenstein:98}
algorithms.  These use only instantaneous particle positions to group
together nearby particles (defined by the FOF `linking length' or the
Denmax `smoothing length') into localized structures. Neither method
performs any type of dynamical analysis to check whether these
structures are bound.

The next level of refinement is to follow a geometrical identification
of halos with a procedure for determining those particles that are
dynamically associated with the substructure. Essentially, we wish to
separate particles belonging to substructure from the local underlying
background matter belonging solely to the cluster.  This we do in two
stages, first removing the velocity outliers from the subhalo, and
then refining the process by iteratively removing the unbound
particles with the greatest total energy until only bound particles
remain. This must be done iteratively to ensure that changes in the
subhalo center of mass and velocity are accounted for as particles are
removed. Examples of publicly available routines that use this
approach include SKID \citep{Stadel:97a, Stadel:01}, BDM
\citep{Klypin:99}, SUBFIND \citep{Springel:01a}, VOBOZ \citep{Neyrinck:04} and MHF \citep{Gill:04a}.  In each of these, the
unbinding is performed assuming the subhalo is completely isolated,
i.e. only the bound particles in the subhalo at each iteration are
taken into consideration when calculating the potential energy.  Not
taken into account in this energy calculation are the (previously
identified) unbound particles located spatially within the subhalo,
nor the disruptive effect of the tidal forces from the particles
surrounding it.  Therefore, in order to correctly identify halos and
subhalos that are not only instantaneously bound, but will also remain
so in the near future (i.e. within the local dynamical timescale), one
should account for all of the forces that may influence the dynamical
state of a subhalo.

An instructive analogy is provided by the stars within the Milky
Way. There are a significant number of stars that, by chance, have an
extremely low velocity relative to that of the Sun. Consider one such
star and the Sun as a single and isolated system, ignoring the effect
of all the other matter in the Milky Way:  by calculating the total
energy of each star, it would appear that they are gravitationally
bound and therefore make up a binary pair. It is only once we also
take into account all the other stars in the Milky Way that we realize
that the gravitational force between the star in question and the Sun
is dwarfed by the tidal forces due to all the other stars in the
galaxy. It then becomes clear that the two stars do not constitute a
stable and bound system. The same argument may be applied to dark
matter substructures in a cluster -- we must take into account all the
forces on a subhalo, including the tidal force from nearby subhalos
and the host halo core, in order to determine whether or not it truly is a
bound system. Hence, previous algorithms, which do not do this,
do not provide a complete picture.

The final refinement to be made, therefore, is to account for all the
remaining forces on each particle in the subhalo. In order to do this
we must now include the effect of the unbound particles on the
potential energy of each subhalo during each iteration of the
unbinding, and then calculate the tidal force on each particle in the
subhalo due to the rest of the particles in the halo. The former is a
simple change to the unbinding algorithm.  Computing the tidal force
on each subhalo particle, however, is a more complex and costly
procedure, especially in higher resolution halos. Since gravitational
forces are linear, we can determine the total force on a particle as
the sum of three terms:
\begin{enumerate}
\item{forces due to the particles in the unit considered,}
\item{forces due to particles within the unit considered, but not bound to that unit,}
\item{forces due to particles outside of the unit considered}.
\end{enumerate}

In this paper we describe a fast algorithm we have developed in order
to approximate the external tidal forces on a substructure.  For the
first time, we present a `coherent' definition of a subhalo that
accounts for all the forces it is subjected to, both internal and
external, and thus identify all particles that are not just
instantaneously bound to the substructure, but will remain so within
the local dynamical timescale. We henceforth refer to the method of
identifying substructures used in previous studies as the {\it
standard} subhalo definition.  Essentially, these use the forces
included in the first term above.  We refer to the new approach
outlined in this study as the {\it coherent} subhalo definition,
allowing for the forces due to all three terms above.  This is for the
purpose of identifying the groups of particles that remain together
over many dynamical timescales, and thus correspond to the luminous
galaxies they host.

In Section \ref{sec:defsam}, we summarize the main steps of the algorithm
common to both procedures (see \citet{Weller:05} for a complete
description), and describe in detail the routine we have developed to
calculate the tidal forces on each particle in a subhalo. We then
demonstrate the impact of each stage of the {\it standard} and {\it
coherent} subhalo definitions on the substructure populations for a
single test halo.  In the last part of this section, we assess the
accuracy of each method by tracking a small sample of subhalos over
several timesteps, checking whether the particles identified as bound
by each definition actually do stay with the subhalo over time. In
Sec. \ref{sec:comp}, we then compare the mass and radial distributions
of substructures for a large sample of halos analyzed using both
definitions. We also look at the mass contained within subhalos of
subhalos -- or sub-substructures -- in each sample to compare the
subhalo populations of successive generations of subhalos in
cluster-mass halos.

\section{DEFINITION OF SAMPLE}\label{sec:defsam}

\subsection{The Identification of Halos and Subhalos}\label{sec:identification}

In our preceding study \citep{Shaw:06}, we presented a comprehensive
analysis of the physical characteristics of a large sample of cluster
mass halos. In order to compare our results with previous work, we
adopted the {\it standard} subhalo definition, as outlined in
\citet{Weller:05}. This algorithm was applied to a $\Lambda$CDM N-body
simulation of $1024^3$ particles with box size $320 h^{-1}\,{\rm
Mpc}$, particle mass ($m_p$) of $2.54\times 10^9 h^{-1} M_{\rm \sun}$ and a spline
kernel force softening length of $\epsilon = 3.2h^{-1}{\rm kpc}$.  The
simulation was evolved to $z$=0.05 using the Tree-Particle-Mesh (TPM)
algorithm \citep{Bode:03a}.  The cosmological parameters used include
$\Omega_m=0.3$, $\Lambda=0.7$, and $\sigma_8$=0.95; outputs from this
run have previously been used to make predictions concerning strong
lensing \citep{Wambsganss:04, Hennawi:05, Das:06}. We wish to ensure
that all the halos are well resolved and that the overmerging problem
is not in evidence.  Thus we discard all halos with a virial mass less
than 10,000 particles ($\approx 3\times 10^{13} h^{-1} M_{\rm \sun}$).

The method described in \citet{Weller:05} starts with the hierarchical
identification of structures. First, cluster mass halos are identified
in the simulation box using a Friends-of-Friends routine with a
linking length of $b=0.2\bar{n}^{-1/3}$, where $\bar{n}^{-1/3}$ is the
mean inter-particle separation. The Denmax routine of
\citet{Bertschinger:91} is then run once on each FOF halo, using a
high resolution smoothing length of $5\epsilon$ in order to identify
the substructures.  A family tree is then constructed by
hierarchically associating the smallest mass subhalos with their
lowest mass `ancestor', so that each subhalo has only a single
immediate parent. Those that consist of less than 30 particles are
dissolved into their immediate parent.

Up until this point the analysis is purely geometrical. Next, we must
refine our identification of subhalos by determining their dynamically
bound constituents. This is achieved in two stages: as a first
approximation we calculate the center of mass velocity of each subhalo
and remove those particles that are statistical outliers \citep[see
Section 2.2.1 in][]{Weller:05}. This step efficiently removes the most
unbound particles in each subhalo, thus allowing a more accurate
determination of the subhalo center of mass velocity. We then complete
the calculation exactly by iteratively identifying those particles
with a total energy greater than zero (in the center of mass frame of
the subhalo) and removing the most energetic. At this point the {\it
standard} and {\it coherent} subhalo definitions diverge. In the
former, once a particle is identified as `unbound' ($E_{bind} > 0$),
it is removed entirely from future iterations of the calculation, and
therefore its contribution to the gravitational potential energy of
the substructure is neglected. This is the procedure that is adopted
in the unbinding step by many publicly available codes.  However, in
the {\it coherent} halo definition, we include the contribution of the
unbound particles within the subhalo.  This has the effect of
increasing the overall binding energy of a subhalo, reducing the
number of unbound particles that are identified. In either approach we
dissolve a subhalo into its immediate parent if at any stage its mass
drops below 30 particles.

In both procedures, we next check to see if pairs of the immediate
daughters of a parent cluster halo are bound. If this is the case,
they form a {\it hyperstructure}; the less massive of the two becomes
a daughter of the more massive structure. If a hyperstructure is
found, we then check to see whether nearby particles previously not
associated with either of the subhalos are bound to it. For the
standard method of analyzing halos, we finally remove all subhalos and
any particles that are not bound to the entire structure and the
procedure is complete. However, in our coherent subhalo definition, we
now allow for the tidal forces on the subhalo due to the external mass
in the cluster; a full description of how this is implemented is
presented in the following section. Once this is done, we finally
remove unbound particles and subhalos from the cluster, as in the
standard analysis.
\begin{figure}
\plotone{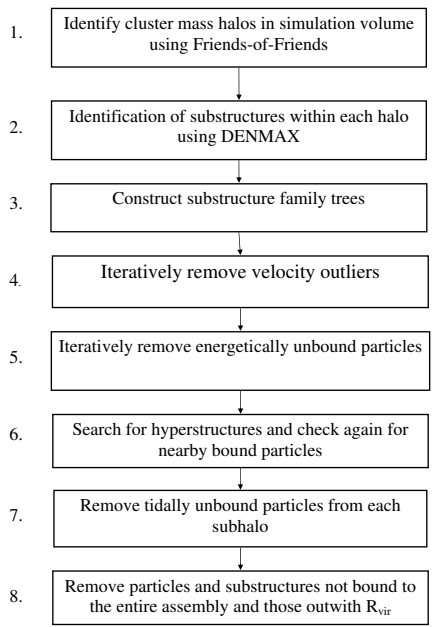}
\caption{Flowchart summarizing the main steps in the identification of
subhalos -- according to our new {\it coherent} subhalo definition -- in
cluster mass halos extracted from dissipationless cosmological N-body simulations.}
\label{fig:flowchart}
\end{figure}

As described in \citet{Weller:05}, both procedures are both stable and
largely independent of arbitrary parameter choices. Henceforth, we
refer to the most massive structure in each cluster as the `mother' or
`host' halo. We will quote the mass of the entire cluster in terms of its
virial mass, $M_{vir}$.  For a $\Lambda$CDM cosmology, it is
conventional to define the virial mass $M_{\rm vir}$ and radius
$R_{\rm vir}$ as $M_{\rm vir}= \frac{4}{3}\pi R_{\rm vir}^3
\Delta_c(z) \rho_c(z)$, where $\rho_c$ is the critical density of the
universe, and the mean over-density $\Delta_c=178\Omega_{\rm
m}(z)^{0.45}$ \citep{Lahav:91}. In order to calculate the virial mass
of each mother halo, we start at its density maximum and proceed
outward until we reach the virial radius, within which the mean
over-density is $\Delta_c$;  we include all substructures with
centers of mass within the virial radius. Hence we define a particle
as being part of the halo if it either belongs solely to the mother
halo or is part of a bound subhalo that has its center of mass within
the virial radius of the mother. A flowchart of all the main steps in the coherent
subhalo definition can be viewed in Fig. \ref{fig:flowchart}.

\subsection{Calculation of external forces: the tidal approximation}\label{sec:tid}

For a subhalo well separated from its mother halo, the tidal radius
$R_t$ is given by \citep{Binney:87a} \beq R_t = r_{\rm d}
\left( {\frac{M(R_t)}{3M(r_{\rm d})}} \right)^{1/3} \, ,
\label{eqn:rtid}
\eeq where $r_{\rm d}$ is the distance between the density peak of the
mother structure and the density peak of the substructure, $M(r_{\rm
d})$ is the mass of all the particles within this radius and $M(R_t)$
is the mass of all the particles within the tidal radius (measured
from the center of the subhalo). This formula simply asks if a
particle feels a tidal force pulling it away from the substructure
larger than the force pulling it to the center of the
substructure. However, this equation does not apply for most halos
under consideration here, since they are well {\em within} the mother
halo. We note that \citet{Kim:06} have recently developed a routine
that uses the tidal radius to demarcate subhalo regions.

Thus, in order to decide if a particle in a daughter halo will be
tidally removed, we must first calculate the tidal force as the
difference between the force on the particle and the force on the
density peak of the daughter: \beq {\mathbf F}_{\rm tid} = {\mathbf
F}_{\rm p} - {\mathbf F}_{\rm d}\, ,
\label{eqn:ftid}
\eeq where for ${\mathbf F}_{\rm p}$ and ${\mathbf F}_{\rm d}$
we sum up the force of all particles from the mother structure,
including all other daughters, outside a radius $R_{\rm t}$.
Particles within the tidal radius are not contributing to the tidal
(i.e.~external) force.  In order to decide if a particle is tidally
bound we calculate the energy 
\beq E_{\rm tid} =
\int_{t}^{t+\tau}{\mathbf F}_{\rm tid}(t^\prime){\mathbf \cdot}
{\mathbf v}_{\rm rel}(t^\prime)\,dt^\prime \approx {\mathbf F}_{\rm
tid}(t){\mathbf \cdot}{\mathbf v}_{\rm rel}(t) \tau \; ,
\label{eqn:etid}
\eeq where $\mathbf v_{\rm rel}$ is the velocity of the particle
relative to the center of mass of the daughter, and $\tau$ is a
characteristic time. As the characteristic time, we use one fifth of
the smaller period of either the particle's orbit around the center of
the daughter halo or the daughter halo's orbit around the center of
the mother; this is the interval over which the tidal force would be
approximately constant. This requirement ensures that there is little
scope for varying $\tau$ as a free parameter. Eqn.~\ref{eqn:etid} thus
gives an approximation for the energy gain of a particle due to the
tidal force in the interval before this force changes
substantially. If a particle has \beq E_{\rm tid}+E_{\rm tot}>0 \; ,
\label{eqn:etid2}
\eeq we declare the particle tidally unbound and move it to the
associated mother. However with this criterion a small number of
particles far from the substructure might be considered bound even
when the tidal force is large, should its velocity by chance be
perpendicular to the tidal force, leading Eqn.~\ref{eqn:etid} to give
a zero tidal energy. We avoid this problem if we include an additional
criterion: \beq {\mathbf F}_{\rm tid} \cdot \left(\frac{\delta
{\mathbf r}}{\left| \delta {\mathbf r}\right|} \right) >
\frac{8}{3}\frac{GM\left(\left|\delta {\mathbf
r}\right|\right)m_p}{\left|\delta {\mathbf r}\right|^2}~,
\label{eqn:ctid}
\eeq where $\delta {\mathbf r}$ is the position of the particle
relative to the central density peak of the daughter, and
$M\left(\left|\delta {\mathbf r}\right|\right)$ is the mass within the
daughter up to this radius. This is a relaxed version of the tidal
radius criterion in Eqn.~\ref{eqn:rtid}, and corresponds to the
particle being at twice the tidal radius if the two halos are
spatially separated.  We declare particles to be tidally unbound if
they fulfil either Eqn.~\ref{eqn:etid2} or Eqn.~\ref{eqn:ctid}.
\begin{figure}
\begin{center}
\includegraphics[scale=.25]{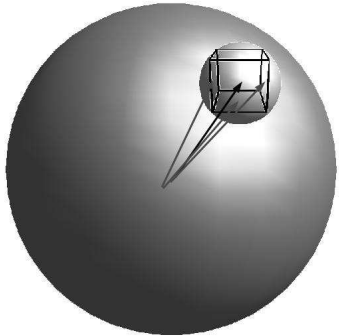}
\caption{Schematic description of the quantities for the calculation
of the tidal radius. The big sphere represents the mother halo, and
the small the daughter. The dark arrow in the middle is the vector to
the density peak of the daughter. The box represents the face of the
cube of length $2 R_t$ which we use to calculate the approximate tidal
forces. The light arrows illustrate three of the positions used to
calculate the forces at the faces of the cube.}
\label{fig:tidal1}
\end{center}
\end{figure}

A problem with the prescribed scheme above is that in order to
calculate the tidal force for each particle we need to perform a
costly $N^2$ operation. However, since an approximate correction for
the tidal force will suffice, we calculate ${\mathbf F}_{\rm tid}$ with a linear
approximation in the following way. We consider a cube with its center
at the density peak of the daughter and a side length of $2 R_{\rm t}$
as in Fig.~\ref{fig:tidal1}.  We calculate the tidal force exactly at
the center of each of the six faces of the cube. We then perform a
linear fit to calculate the tidal force at the center of the each of the faces with 
\beq {\mathbf F}_{\rm tid} = {\mathbf A} \left({\mathbf r}_{\rm p}-{\mathbf
r}_{\rm d}\right)\, ,
\label{eqn:ftidapprox}
\eeq where ${\mathbf r}_{\rm d}$ is the peak density position of the
daughter (relative to the peak density position of the mother halo),
and the entries of the $3\times 3$ matrix {\bf A} are obtained by a
least squares fit of the exact tidal force at the center of the 6
faces.  This approximation corresponds to a divergence free field and
hence to a quadrupole approximation, and works well when the tidal
radius of the daughter is sufficiently small compared to the distance
between the density peak of the daughter and the mother.  We use the
approximation if $R_{\rm t}/ \left|{\mathbf r}_{\rm d}\right| <0.5$;
otherwise we switch to an exact $N^2$ calculation.  We found that the
median error due to the linear tidal force approximation was about
20\%.
\begin{figure}
\begin{center}
\includegraphics[scale=.25]{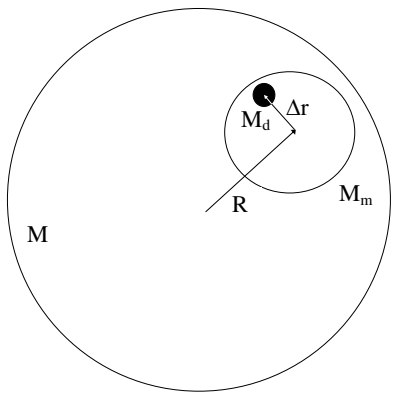}
\caption{Schematic representation of the quantities required to
  establish if a daughter is tidally bound.}
\label{fig:tiddaughter}
\end{center}
\end{figure}

The next step is to remove tidally unbound daughters from
structures. From the point of view taken here, none of the daughters
can be tidally removed from the entire cluster, because we are unable
to calculate the tidal forces from the mass distribution surrounding
it. So this step will not change the total amount of substructure in
the halo, but it might reduce the size of hyperstructures.  The
criterion to determine if a daughter is tidally bound to a another
substructure is \beq \Delta r \le
R\left(\frac{m_d+m_M}{3M}\right)^{1/3}~, \eeq with $M$ the mass of the
mother of the substructure, $m_M$ the mass of the substructure, $m_d$
the mass of the daughter of the substructure, $R$ the distance from
the mother to the density peak of the substructure, and $\Delta r$ the
distance between the substructure and the density peak of the
daughter. These quantities are schematically represented in Figure
\ref{fig:tiddaughter}.

\subsection{Application to a Single Halo}

In order to compare the standard and coherent halo definitions, in
Table \ref{tab:steps} we give for each step the number of particles
($N_p$), the number of subhalos and the fraction of mass contained in
substructure ($f_s$) for one of the most massive halos in our sample,
analyzed using both methods. The first line in the table gives the
properties of the halo at the point just prior to where the two
methods diverge -- after the initial geometrical identification of
subhalos using DENMAX and having removed the velocity outliers from
each subhalo \citep[see ][]{Weller:05}. During the following two
stages -- the unbinding and identification of hyperstructures -- over
half the substructure is removed in both methods. However, at this
point there is nearly $77\%$ more substructure in the halo with the
coherent analysis relative to the standard. This is because in the
coherent analysis the unbound particles are retained in the potential
calculation (but ignored in the standard analysis).  Therefore, the
gravitational potential experienced by each particle is greater, thus
decreasing the number of particles identified as being
`unbound'. However, once the tidally unbound particles have been
removed, the fraction of mass contained in substructure is only
slightly greater for the coherent analysis. $N_p$, $f_s$ and the
number of subhalos all decrease further in the final stage as
particles and subhalos not bound to the entire cluster, or those that
are outwith the virial radius, are removed (see the last line of Table
\ref{tab:steps}).

\begin{center}
\begin{figure}
\epsscale{0.6}
\centerline{\plotone{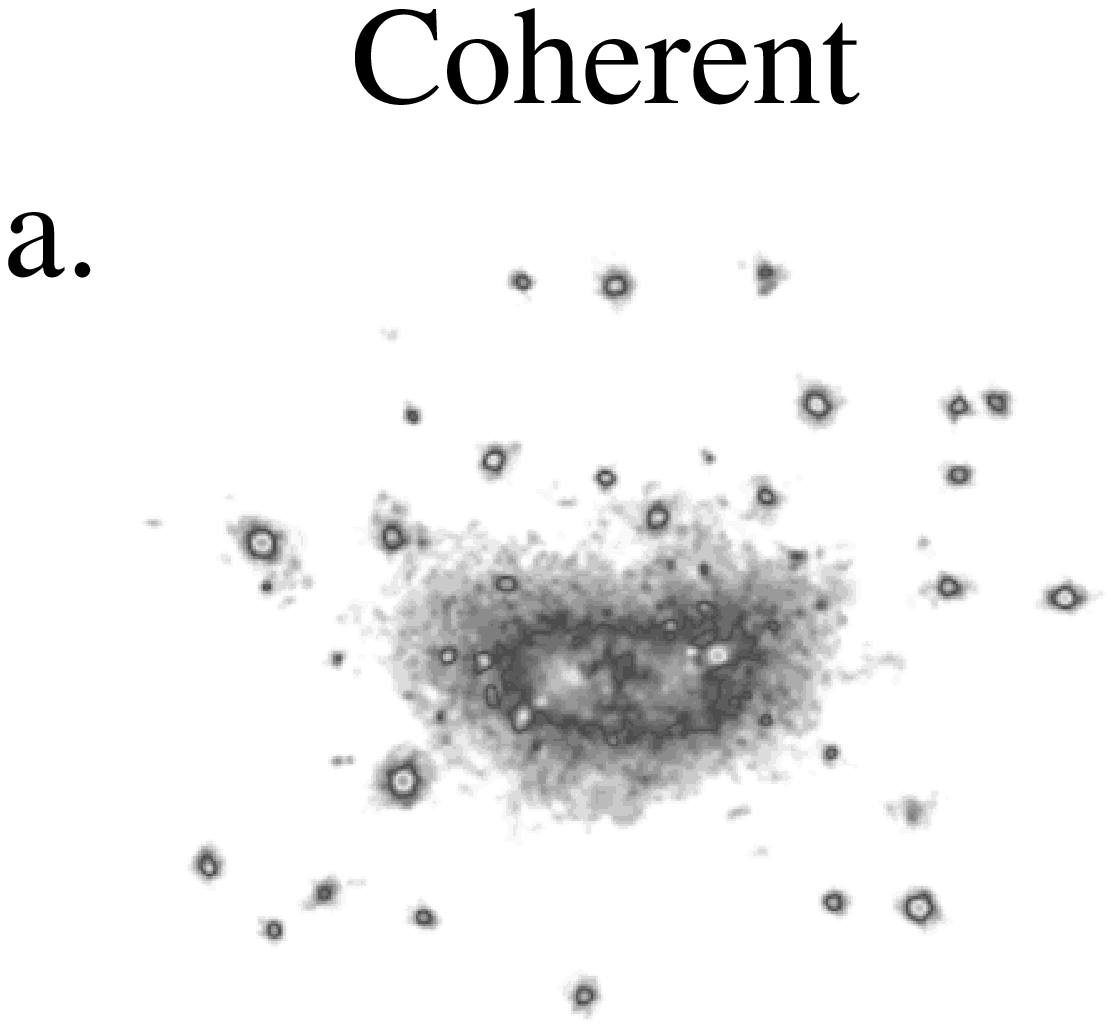}\plotone{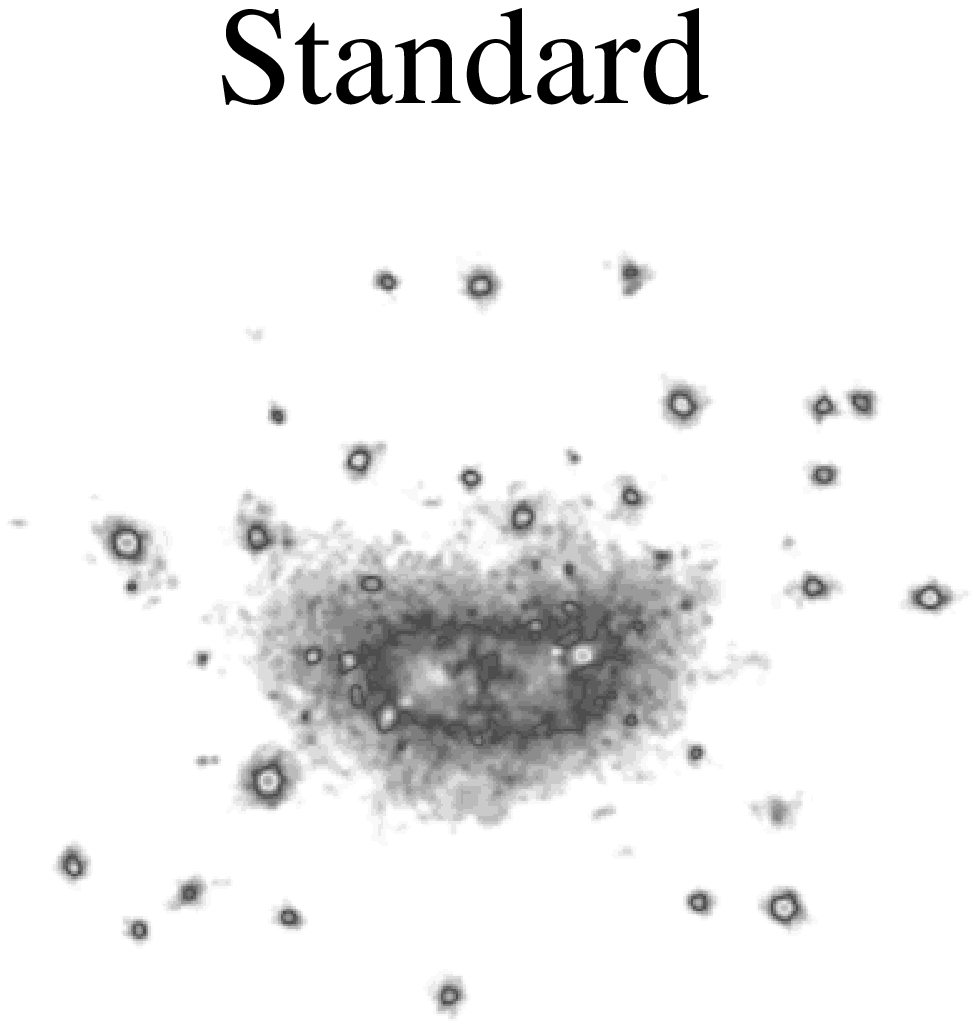}}
\vspace{-0.5cm}
\centerline{\plotone{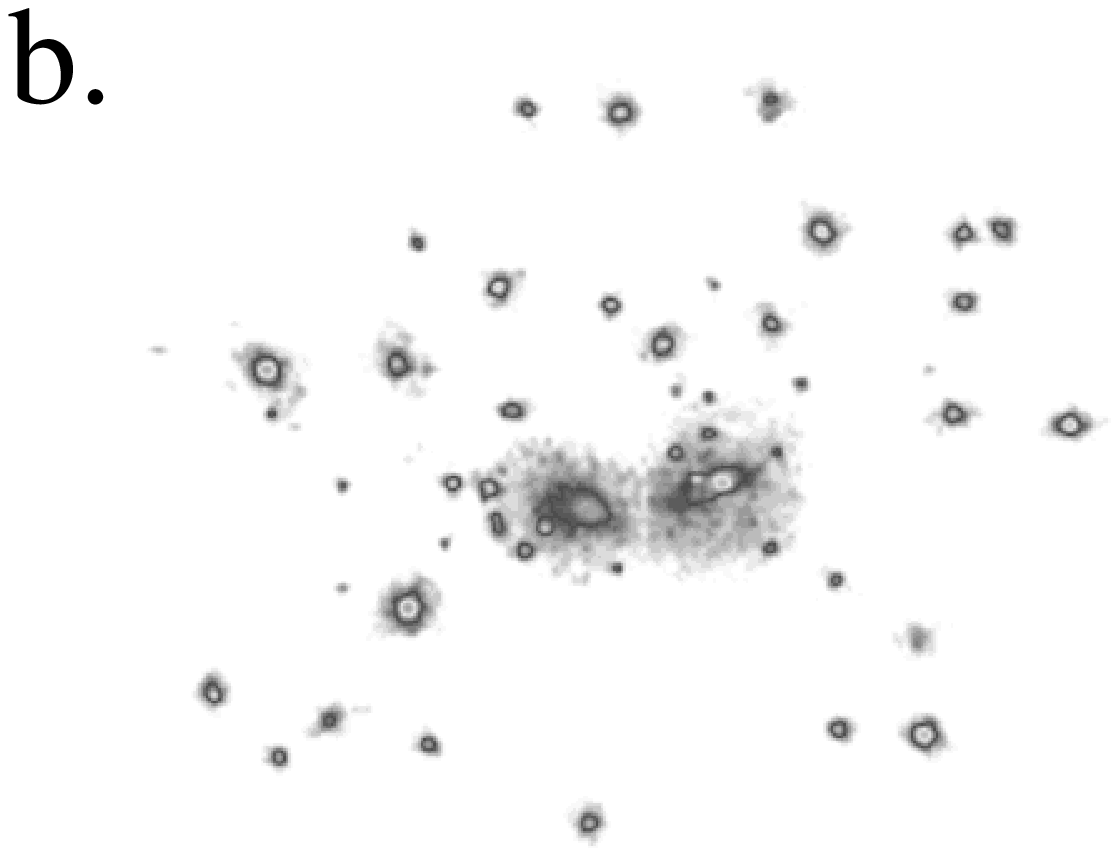}\plotone{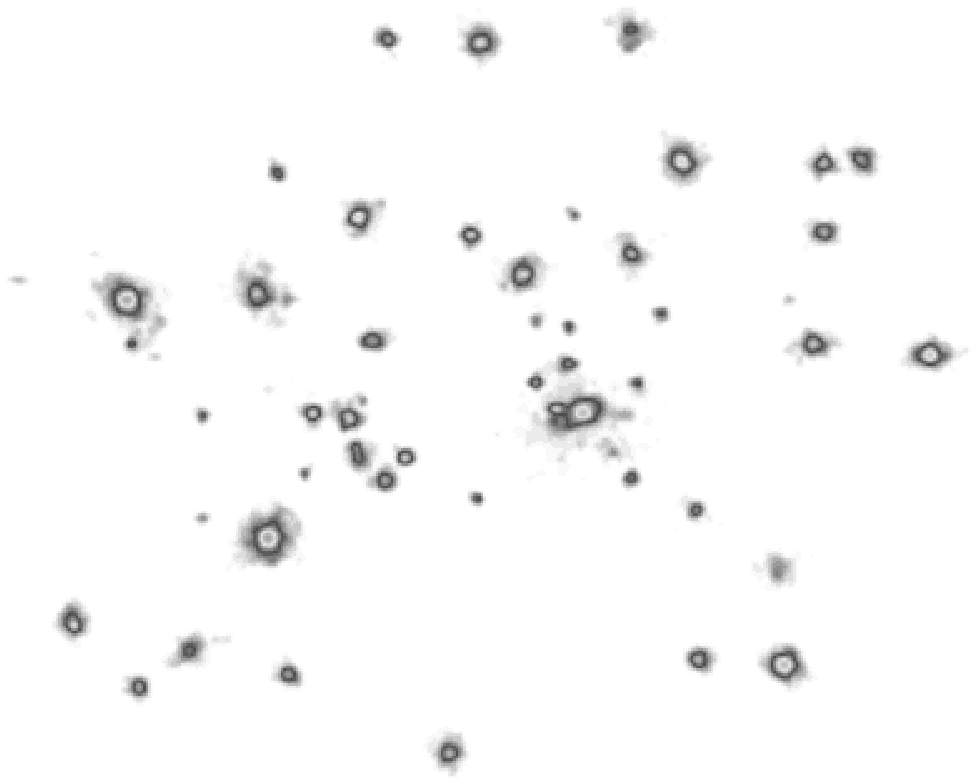}}
\vspace{-0.5cm}
\centerline{\plotone{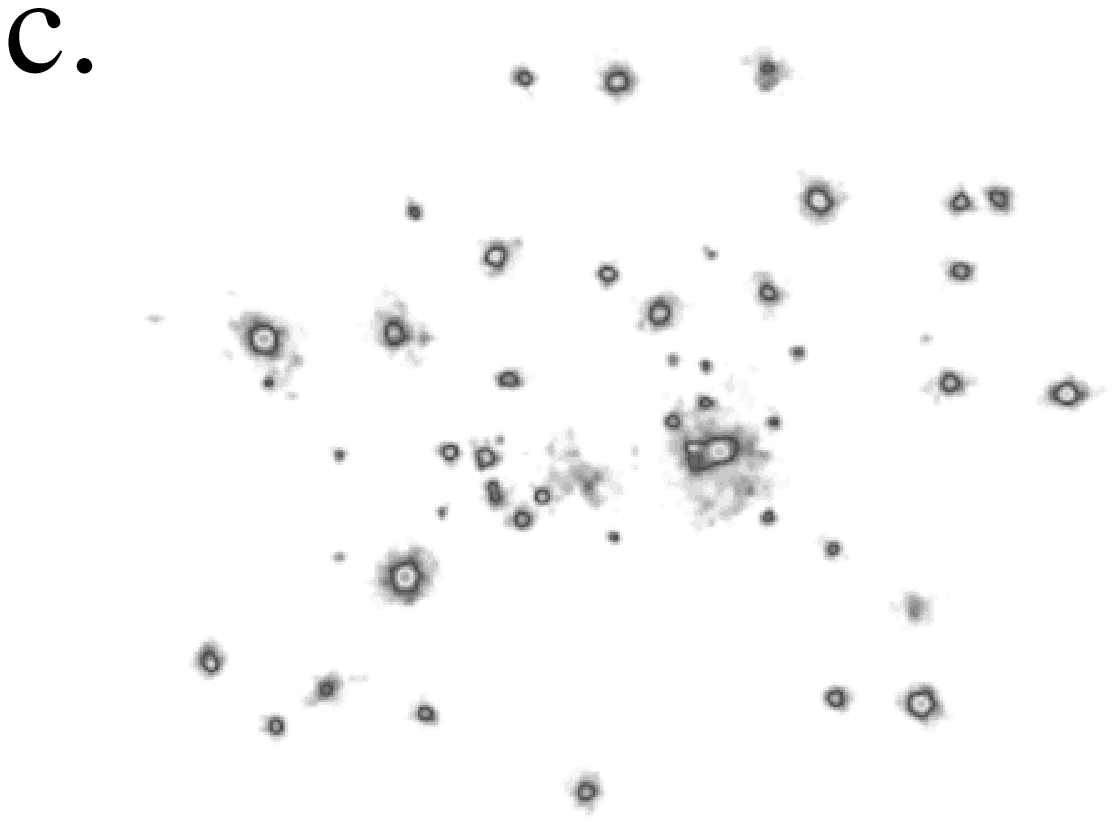}\hspace{4.5cm}}
\vspace{-0.5cm}
\centerline{\plotone{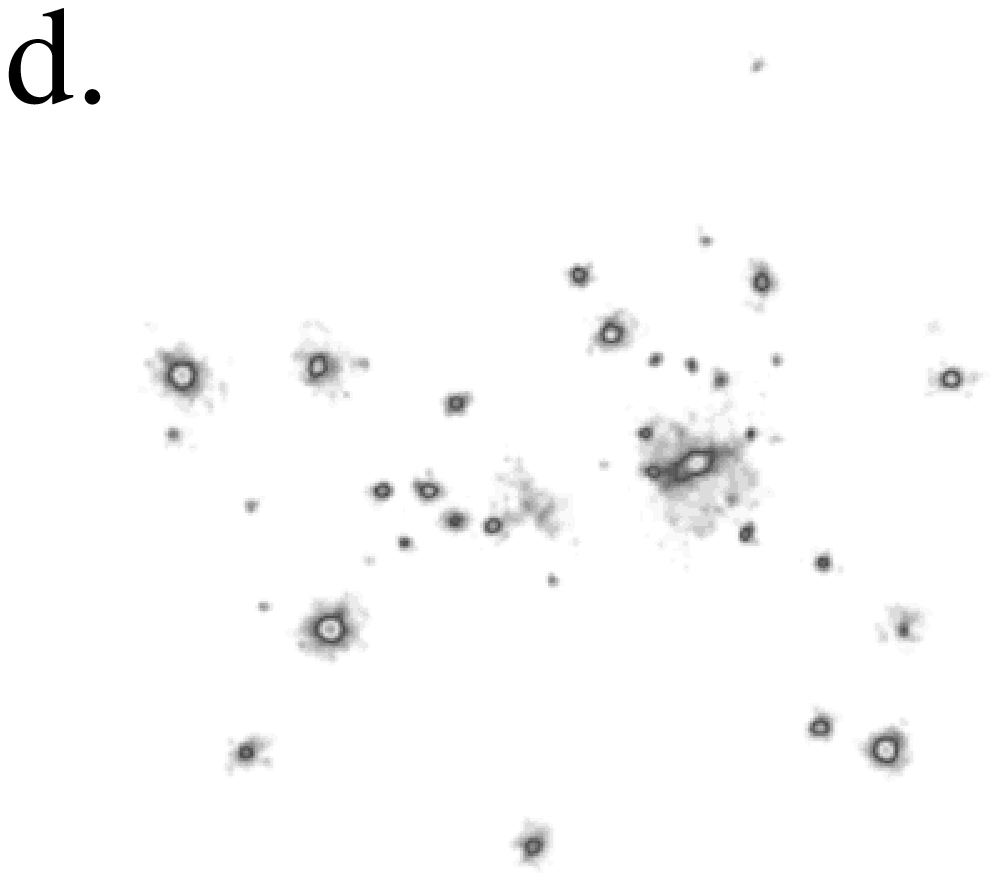}\plotone{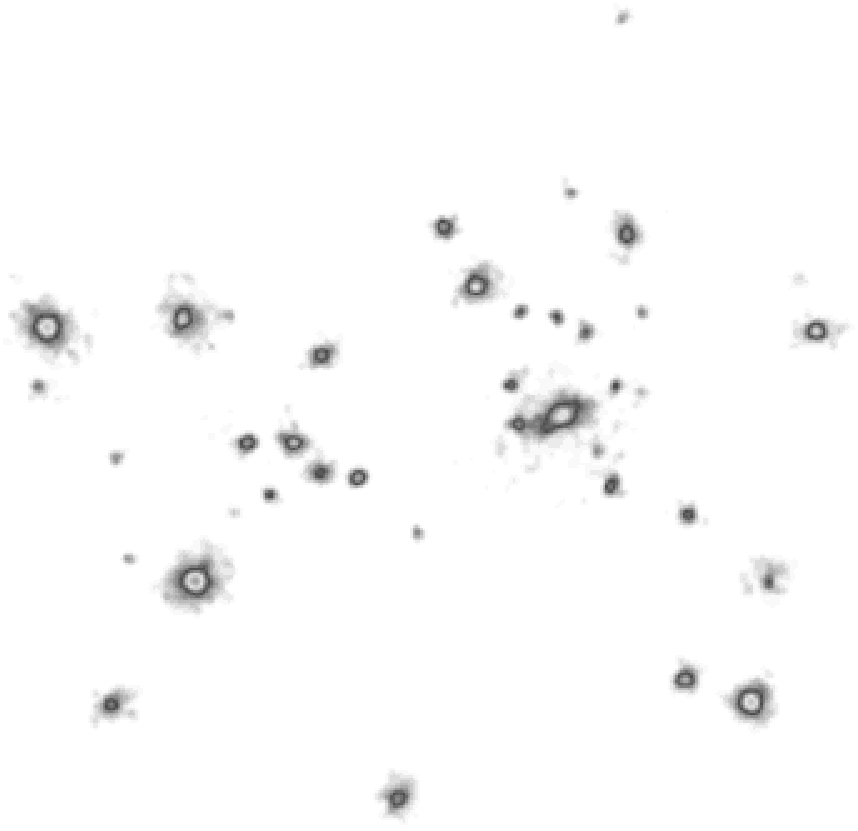}}
\vspace{-0.5cm}
\caption{2d projection of the substructure in one of the
most massive halos in our sample after the completion of each stage of
the analysis for both the {\it standard} and {\it coherent} halo
definitions: after the denmax step and removal of the velocity
outliers (a), the unbinding and identification of hyperstructures (b),
the removal of tidally inbound particles (c) and, finally, having
removed particles and subhalos unbound to the entire structure (d).}
\label{fig:halo_pics}
\epsscale{1.0}
\end{figure}
\end{center}

Fig. \ref{fig:halo_pics} displays projections of the density of
substructure particles in the halo after each step in Table
\ref{tab:steps}. The greater mass of subhalos after the unbinding step
in the coherent analysis relative to the standard analysis is evident,
as is the impact of the tidal step.
\input{tab1.tex}
\subsection{Tracking Subhalo Mass}\label{sec:trackhalos}

We have described a new method for defining and measuring the bound
mass of substructures in N-body simulations. However, in order to
demonstrate decisively that our `coherent' scheme is an improvement on
the standard procedure, we must show that the particles it identifies
as `bound' are still associated with the subhalo at a later time, and
that it is more accurate in doing so. Likewise, we must demonstrate
that the particles identified as `unbound' are not associated with the
subhalo at future timesteps.

To perform this test, we use the output of a $256^3$ particle
simulation (with particle mass $m_{p} = 3.15\times 10^7 \; h^{-1} \; M_{\odot}$)
in a box of side-length $19.25 \; h^{-1} \; Mpc$ and a (spline)
softening length of $2.5 h^{-1} kpc$. We adopt a $\Lambda$CDM
cosmology, with $\Omega_{\Lambda} = 0.733$, $H_0 = 71 \; km s^{-1}
Mpc^{-1}$ and $\sigma_8 = 0.84$. The output of this simulation was
saved after every 10 particle-mesh timesteps (see Table
\ref{tab:tsteps});  henceforth, we refer to the time elapsed between
each saved output as being one `timestep'. Each particle was allocated a
unique identifier number at the beginning of the simulation, to enable
us to track individual particles over time.  Although this simulation
contained fewer particles that our main $1024^3$ simulation, the
smaller box size ensured that we could easily locate and follow a
small number of well resolved cluster-mass halos.

In each of the final six outputs of the simulation (from z = 0.13 to
0, or a time period of $1.17$ Gyrs) we identified all structures of
mass greater then $10^{13} M_{\odot}$ using the procedure outlined in
Sec. \ref{sec:identification}. From this sample we selected the most
massive halo possessing a stable mass -- one in which the mass
fluctuates by less then 5\% -- over the time-period considered. The
main physical properties of this halo can be viewed in Table
\ref{tab:tsteps}. We also checked to make sure that the halo did not
undergo any major mergers during this period.

At each timestep, we applied both the standard and coherent subhalo
definitions to identify the substructure populations of the halo. For
each subhalo we identify five sets of particles. The set $D_i$
contains all the particles geometrically associated with a subhalo, as
identified by the DENMAX step (step 2 in Figure \ref{fig:flowchart}),
at time $t_i$ (where the index i indicates the timestep) . The subset
$U_i$ contains all the particles that were associated with a subhalo
by the DENMAX routine (i.e. that are in $D_i$) but were then removed
during the unbinding stages. The subset $B_i$ contains all particles
that remained in the subhalo at the end of our entire routine -- these
are the `bound' particles.  Hence, for each subhalo we have five sets
at each timestep; $D_i$, which is of course defined before the binding
criteria are applied and so is the same for both the standard and
coherent definitions, $U_i$ and $B_i$ as determined by the standard
definition and $U_i$ and $B_i$ as determined by the coherent. The
number of particles in any set $S_i$ is written N($S_i$). For example,
the total number of particles associated with a subhalo at a
particular timestep is ${\rm N}(D_i) = {\rm N}(U_i) + {\rm N}(B_i)$.

To achieve our aim of following the bound constituents of a subhalo,
we must be able to identify the same subhalo at a later time. This is
more complicated then simply matching up particle id tags; over the
course of time, subhalos may dissolve (by dropping below the 30
particle limit), merge, fragment, or temporarily move outwith the
virial radius of the cluster or disappear into its core where DENMAX
is unable to locate them. Hence, to achieve a one-to-one
correspondence between subhalos at two different times requires a
clear definition of how to identify its descendant. To resolve this
issue we developed a simple matching algorithm that identifies the
descendant of a subhalo when there might be some ambiguity. This
routine operates by pairing the most massive progenitor with its most
massive descendant, should there be more than one in either case.
\input{tab2.tex}

At each timestep considered, $t_i$, we look for the same subhalo at
the subsequent timestep $t_{i+1}$. We do this for both the standard
and coherent methods. If no descendant at the subsequent timestep is
found for a particular subhalo for either method it is removed from
the sample. This ensures a consistent comparison between the two
methods (i.e. the same subhalos are being compared). Normally, a
subhalo will only permanently disappear if it drops below the 30
particle minimum mass and is dissolved into its immediate parent. For
each subhalo, we are now able to calculate quantities such as ${\rm
N}(B_i \cap B_{i+1})$, or the total number of particles that are
found to be bound to the subhalo at {\it both} $t_i$ and $t_{i+1}$.

We now define three quantities (to be measured for each subhalo) that
can be used to assess the accuracy of each criterion:

\begin{itemize}
\item{The probability $P_b$ that a particle in the subhalo is contained
in the `bound' set $B_i$ but is {\it not} in the `bound' set $B_{i+1}$
at the subsequent timestep (written $\bar{B}_{i+1}$):
\begin{equation}
P_b = \frac{{\rm N}(B_i \cap (\bar{B}_{i+1}))}{{\rm N}(D_i)} \;.
\end{equation}
Note that $\bar{B}_{i+1} \neq U_{i+1}$. This is because $U_{i+1}$
includes only those particles that are geometrically associated to a
subhalo at $t_{i+1}$, but were removed from the subhalo during the
unbinding stages. $\bar{B}_{i+1}$ includes any particle in the entire
{\it halo} that is not bound to the subhalo in question at $t_{i+1}$.}
\item{The probability $P_u$ that a particle in the subhalo is
contained in the `unbound' set $U_i$ but is found in the `bound' set
$B_{i+1}$ at the subsequent timestep:
\begin{equation}
P_u = \frac{{\rm N}(U_i \cap B_{i+1})}{{\rm N}(D_i)} \;,
\end{equation}
}
\item{The error on the bound mass determination for a subhalo,
\begin{equation}
\epsilon_m = \frac{{\rm N}(B_i) - {\rm N}(B_{i+1}|D_i)}{{\rm N}(D_i)} \;,
\end{equation}
} or in words, the fraction of particles that were labelled `bound' at
$t_i$, minus the fraction of particles from the total set $D_i$ that
actually {\it were} bound (i.e. are also found in $B_{i+1}$). 
\end{itemize}
The reason why we must write $(B_{i+1}|D_i)$ is that subhalos
sometimes `sweep up' particles from the mother halo between
timesteps. Of course, these particles cannot be accounted for by
either criteria -- we therefore ignore them in this analysis. 

If a criterion for identifying the bound mass component of subhalos in
simulations claimed to operate perfectly, we might expect to find $P_b
= 0$ (particles that we say are bound, do actually stay bound), $P_u =
0$ (particles that we say are unbound move away) and $\epsilon_m = 0$
(the number of particles identified as bound at $t_i$ is the same as
the number identified as bound at $t_{i+1}$, ignoring particles picked
up along the way). In practice, within one timestep a subhalo may
undergo a close encounter with another subhalo or the host halo core
and therefore lose particles due to physical processes that cannot be
accounted for by applying dynamical criteria at each
timestep. However, these effects will apply equally to subhalos
identified by both methods, and thus will not affect our study of
their relative accuracy.

\begin{figure}
\plotone{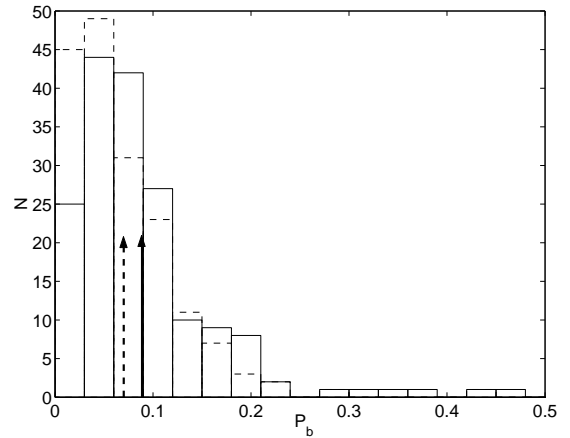}
\caption{The distribution of $P_b$ (see Section \ref{sec:trackhalos})
for the coherent (dashed) and standard (solid) subhalo samples,
measured over a single timestep. The arrows mark the mean values of
each distribution, respectively. It is clear that the coherent method
is less likely to mislabel as `bound' particles that infact leave the
subhalo.}
\label{fig:eb}
\end{figure}

\begin{figure}
\plotone{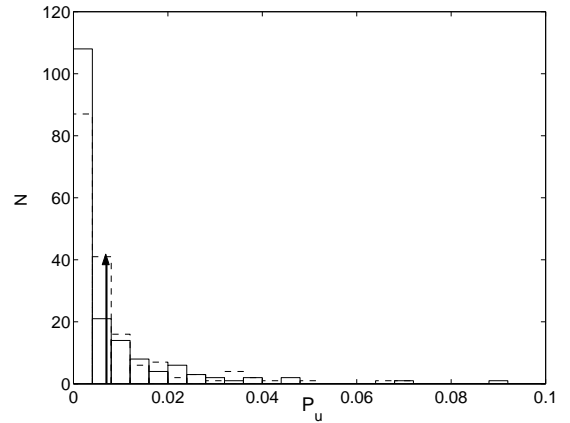}
\caption{The distribution of $P_u$ (see Section \ref{sec:trackhalos})
for the coherent (dashed) and standard (solid) subhalo samples,
measured over a single timestep. The arrow marks the mean values
(which are the same) of both distributions. On average, the results
indicate that, for both methods, less then 1\% of the particles
identified as `unbound' are actually bound to (and thus remain with)
the subhalo.}
\label{fig:eu}
\end{figure}

\begin{figure}
\plotone{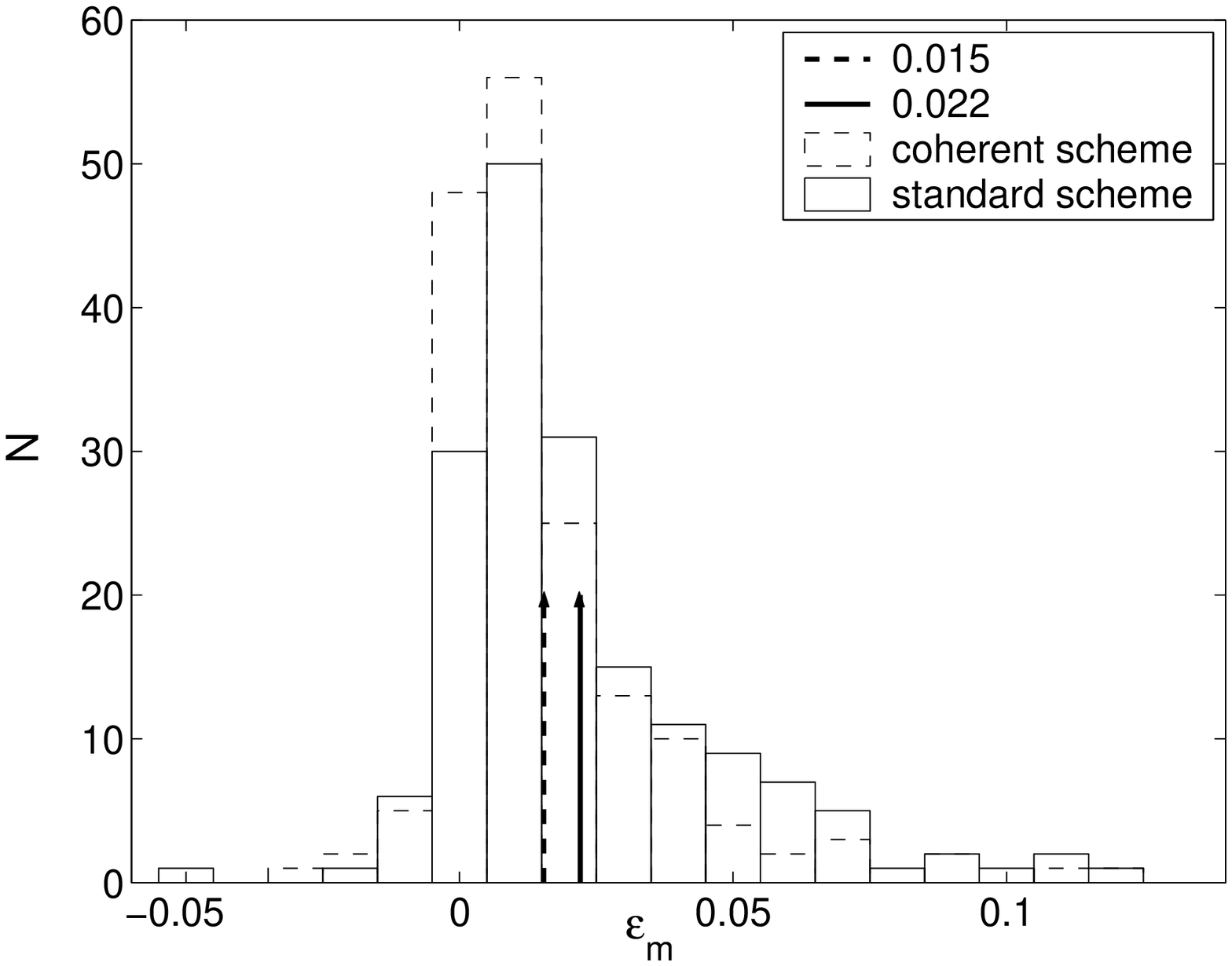}
\caption{The distribution of $\epsilon_m$ -- the error on the bound
mass determination for each subhalo -- (see Section
\ref{sec:trackhalos}) for the coherent (dashed) and standard (solid)
subhalo samples, measured over a single timestep. The arrows mark the
mean values of each distribution, respectively. The coherent method is
clearly more accurate at measuring the bound mass of a subhalo than
the standard method.}
\label{fig:em}
\end{figure}

In Figures \ref{fig:eb}, \ref{fig:eu} and \ref{fig:em} we plot the
distribution of each of these quantities for the full sample of
subhalos followed over a single timestep (i.e. i = 35-36, 36-37,
etc). This is equivalent to 0.23 Gyrs, or 2.88 times the average
characteristic dynamical time ($\tau$ in Section \ref{sec:tid}). It is
clear from the distributions of $P_b$ in Figure \ref{fig:eb} that the
coherent definition (dashed line) is less likely to mislabel as
`bound' a particle that does not remain with a subhalo than the
standard definition (solid line). The mean value of $P_b$ for the
standard sample ($0.089 \pm 0.006$, where the error is the standard
error in the mean) is 28\% greater than that for the coherent sample
($0.070 \pm 0.004$). Hence, the standard routine is less successful at
removing particles that are not bound to the subhalo.  However, both
methods do equally well at correctly identifying the unbound particles
-- the mean value of both distributions in Figure \ref{fig:eu} is
$\langle P_u \rangle = 0.007$.


The distributions of $\epsilon_m$ in Figure \ref{fig:em}, which are
largely greater then zero, suggest that both methods overestimate the
bound mass of a subhalo. However, it is clear that the error is
significantly less for the coherent definition ($\langle \epsilon_m
\rangle = 0.015 \pm 0.002$, dashed line/arrow) than the standard
subhalo sample ($\langle \epsilon_m \rangle = 0.022 \pm 0.002$, solid
line/arrow). Again, this demonstrates that the standard routine is
less accurate at identifying and removing unbound particles and thus
provides a less reliable measure of subhalo mass.

\begin{figure}
\plotone{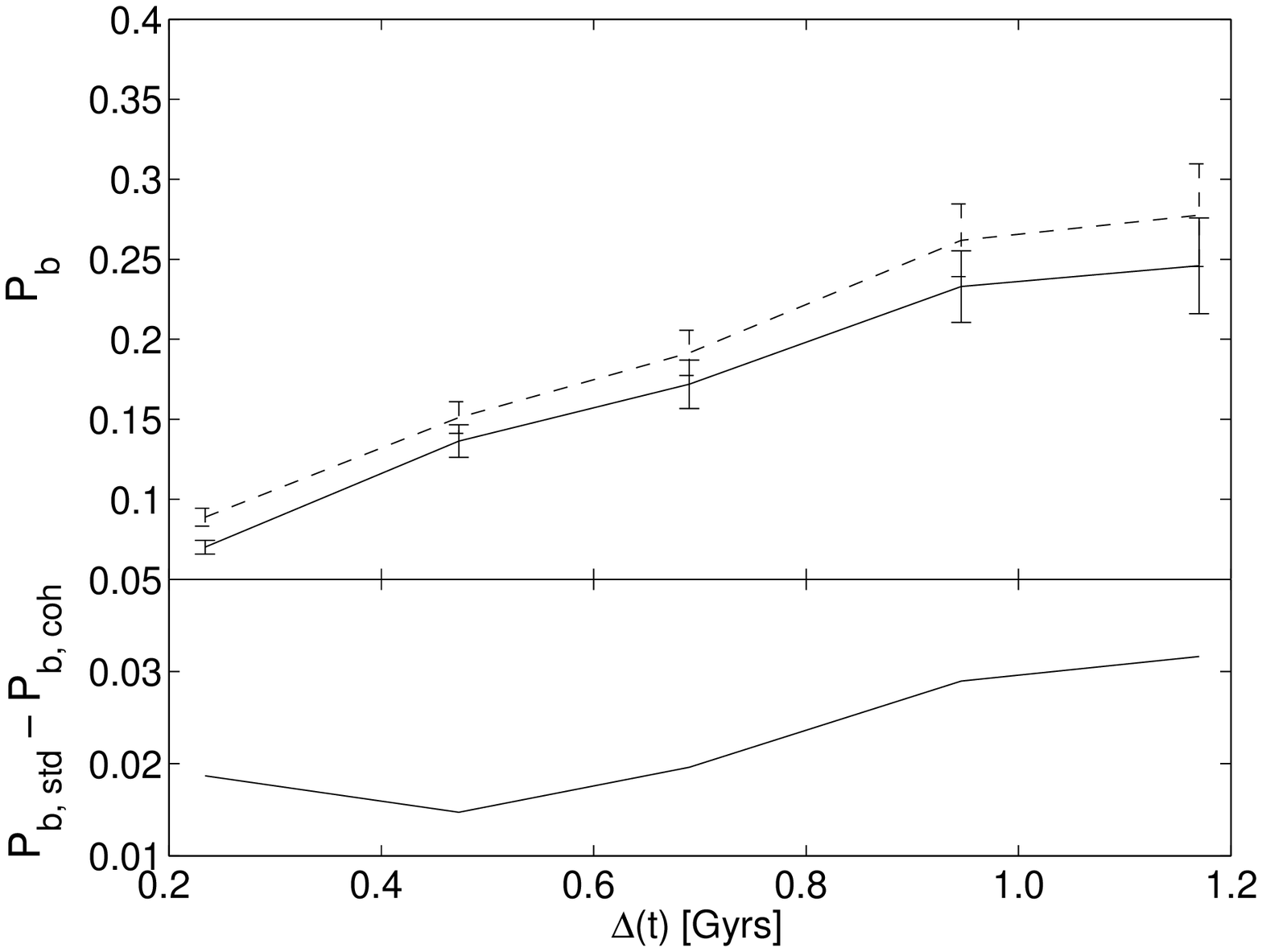}
\caption{({\it upper}) The mean value of $P_b$ obtained when measured
over 1-5 timesteps ($\Delta t$ = 0.23 - 1.17 Gyrs, or 2.88 - 14.6
characteristic dynamical times) for the standard (solid) and coherent
(dashed) subhalo definitions. Error bars denote the standard error in
the mean. ({\it lower}) The difference between the mean values of
$P_b$ measured for the standard (std) and coherent (coh) methods as
$\Delta t$ increases.}
\label{fig:eb_t}
\end{figure}

\begin{figure}
\plotone{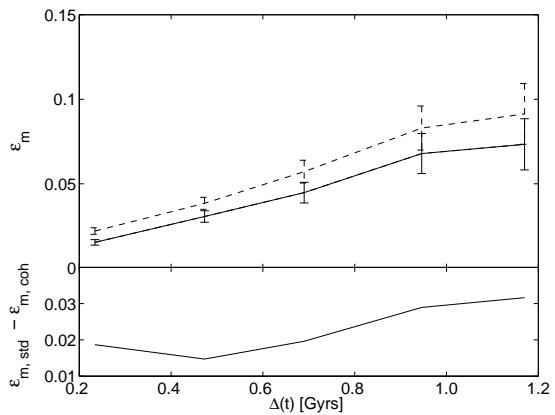}
\caption{({\it upper}) The mean value of $\epsilon_m$ obtained when
measured over 1-5 timesteps ($\Delta t$ = 0.23 - 1.17 Gyrs, or 2.88 -
14.6 characteristic dynamical times), for the standard (solid) and
coherent (dashed) subhalo definitions. Error bars denote the standard
error in the mean. ({\it lower}) The difference between the mean
values of $\epsilon_m$ measured for the standard (std) and coherent
(coh) methods as $\Delta t$ increases.}
\label{fig:em_t}
\end{figure}

We now investigate the accuracy of each method as a function of time
(number of timesteps). This is achieved by also matching subhalos to
their descendants between two and five timesteps later (equivalent to
$~0.473 - ~1.17$ Gyrs, or 2.88 - 14.6 characteristic dynamical times)
and determining the fraction of particles in each subhalo at $t_i$
that were correctly identified as being bound or unbound. In Figures
\ref{fig:eb_t} and \ref{fig:em_t} we analyse $\epsilon_m$ and $P_b$ as
a function of the time between outputs ($\Delta t$). In the upper
panel of both plots, each point represents the mean values (and the
standard error in the mean) of $\epsilon_m$ and $P_b$ respectively,
for the sample of subhalos followed over 1-5 timesteps. Hence, the
left-most point in the upper panel of Figure \ref{fig:eb_t} represents
the mean value of $P_b$ measured over a single timestep (as marked by
the arrows in Figure \ref{fig:eb}), the second point over two
timesteps, and so on. As usual, the dashed line represents the results
obtained for the coherent sample; the solid line those obtained for
the standard sample.

As expected, $P_b$, the probability that a particle labelled as
`bound' at time $t_i$ has left the subhalo at a later time $t_{i+j}$
(where j=1-5), increases as $\Delta t$ increases for both methods (see
Figure \ref{fig:eb_t}). After a single timestep ($\approx 0.23$ Gyrs),
$P_b = 0.07$ and $0.09$ for the coherent and standard methods
respectively. When measured over five timesteps ($1.17$ Gyrs), this
increases to $0.25$ and $0.28$. The lower panel, which charts the
difference in the value of $P_b$ obtained between the standard and
coherent method, demonstrates that the coherent procedure becomes
increasingly more accurate with respect to the standard with larger
$\Delta t$. A similar result is demonstrated in Figure \ref{fig:em_t},
where we plot $\epsilon(\Delta t)$. Gradually, particles identified as
`bound' by each method are removed from the subhalo by dynamical
friction or tidal stripping, and so the error on the predicted subhalo
mass at $t_i$ grows with $\Delta t$. However, as discussed above, the
increase in $\epsilon$ and $P_b$ will be partly due to interactions
with other subhalos or the cluster core within the time period over
which they are evaluated and therefore is not solely due to
inaccuracies in either method. Nevertheless, the errors for the
standard method grow slightly more rapidly than for the coherent.

Overall, although both methods do well at determining those particles
that are bound to a subhalo and those that will soon move away, we
have clearly demonstrated that our new coherent method provides the
more precise measure of subhalo mass. Whilst both the standard and
coherent definitions overestimate subhalo masses, the latter is less
prone to mistakenly labelling `unbound' particles as `bound'. It is
important to note that the improvement is evident despite the 20\%
error in the tidal forces due to the approximate scheme that we use to
evaluate them (see Section \ref{sec:tid}).  Additionally, at any
instant, we are only able to determine those particles that will be
removed by tidal forces within a single characteristic dynamical
timescale ($\langle \tau \rangle \approx 0.08$ Gyrs = 0.35 timesteps),
as described in Section \ref{sec:tid}. We therefore do not expect our
coherent definition to be significantly more accurate then the
standard scheme over timescales longer than a few dynamical times.

Furthermore, to perform this analysis we picked a halo that contained
a typical fraction of its mass in substructure ($\approx$ 7\%). If we
had chosen a halo that contained several large substructures (i.e. a
halo undergoing a major merger) then the tidal forces exerted on each
subhalo would have been more substantial, requiring a larger
correction to their bound mass.

\section{Comparison between methods for a large halo sample}\label{sec:comp}

Both the standard and coherent subhalo definitions were applied to the
2000 most massive halos extracted from the large $1024^3$ particle
simulation described in Section \ref{sec:identification}. Using the
procedure described in \citet{Shaw:06} we removed from each sample
those halos that are not `virialized', that is, have not yet reached a
state of dynamical equilibrium; there then remains 1838 halos in each
sample. In the following Section we compare the subhalo populations
and their radial distributions for each sample, and finally we
investigate the subhalo populations of {\it subhalos}.

\subsection{Substructure Populations of Halos}

Since the force and mass resolution of N-body simulations has reached
a high enough level to enable the identification of subhalos
within larger structures, several previous studies have attempted to
measure the slope of the subhalo mass function (SMF). This is
especially important if one wishes to quantify the relationship
between the properties of dark matter subhalos identified in
simulations and observed properties of galaxies 
\citep[see for example][]{Kravtsov:04b, Tasitsiomi:04b, Vale:06}. 
Studies of small numbers
of halos selected from low resolution simulations and resimulated at
higher resolutions have suggested that dark matter halos are
self-similar in terms of their subhalo populations: low mass halos
appear like `rescaled versions' of higher mass halos \citep{Moore:99a,
Ghigna:00, DeLucia:04, Reed:05}. However, recent studies of much
larger samples have found evidence that this is not the case -- higher
mass halos contain more substructure, as they have formed more recently
and have therefore had less time in which to disrupt their hosted
subhalos \citep{Gao:04, Kang:05, Shaw:06}.

\begin{figure}
\plotone{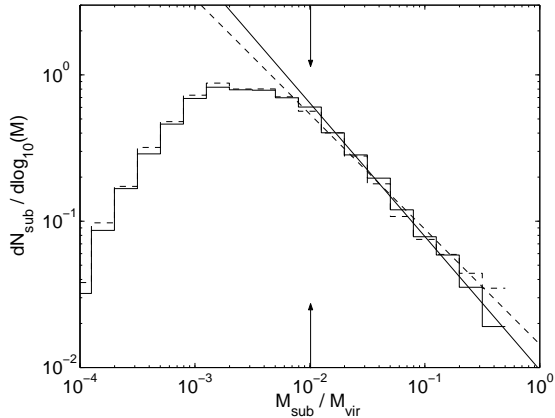}
\caption{Mass distribution of all the subhalos in our coherent (dashed
lines) and standard (solid) halo samples as a function of the ratio of
subhalo to cluster halo mass. Straight lines are the power law fits to
the slope outwards of the point indicated by the arrow, which is where
we assume that the distribution is not affected by the minimum subhalo
mass in our simulation (30 particles, or $7.62\times 10^{10} h^{-1}
 M_{\rm \sun}$). Slopes of the fit are $-0.79\pm 0.04$ and $-0.91\pm 0.03$ for
the coherent and standard distributions respectively.}
\label{fig:mass_func}
\end{figure}

In Fig. \ref{fig:mass_func} we compare the differential subhalo mass
functions for each of our halos samples, quoting subhalo masses as a
fraction of the cluster virial mass. The flattening off and eventual
decline of each distribution for decreasing subhalo masses is due to
the minimum substructure mass in our simulation ($7.62\times 10^{11}
h^{-1}  M_{\rm \sun}$, or 30 particles). For subhalos with $M_{sub} > 0.01
M_{vir}$ this effect appears not to be in evidence. Therefore, for
subhalo masses greater than this, we fit a power-law --
$dN/dlog_{10}(M) = (M_{sub} / M_{vir})^{\alpha}$ -- to the
distributions, obtaining a slope of $-0.79 \pm 0.04$ for the coherent
sample (dashed lines) and $-0.91 \pm 0.03$ for the standard halo
sample (solid lines). Typically, most published studies have found
that a power-law fit to the slope of the subhalo mass function results
in values of $\alpha$ between -0.7 and -1.1 \citep{Moore:99a,
Ghigna:00, Helmi:02, Gao:04, DeLucia:04, Shaw:06, vandenBosch:05}. Our
results are therefore well within the range measured by other authors.

In general, there is good agreement between the two halo samples,
although there is a greater number of very high mass subhalos
($M_{sub} > 0.2M_{vir}$) in the coherent sample, resulting in a
shallower slope than measured for the standard sample. On average,
these high mass subhalos tend to reside within slightly lower mass
clusters ($\langle M_{vir} \rangle \approx 1.5 \times 10^{14} M_{\rm
\sun}$) than for the entire sample ($\langle M_{vir} \rangle \approx 3\times
10^{14} M_{\rm \sun}$). They are also typically located near to the
cluster core, where the density of background particles -- those bound
only to the mother halo -- is greater.  In the coherent subhalo sample
these halos will tend to be more massive than their counterparts in
the standard sample, as we include the contribution of the background
particles to the binding energy of a subhalo.  Furthermore, they are
not subjected to strong tidal forces from the halo core because they
are located in lower mass halos.
\begin{figure}
\plotone{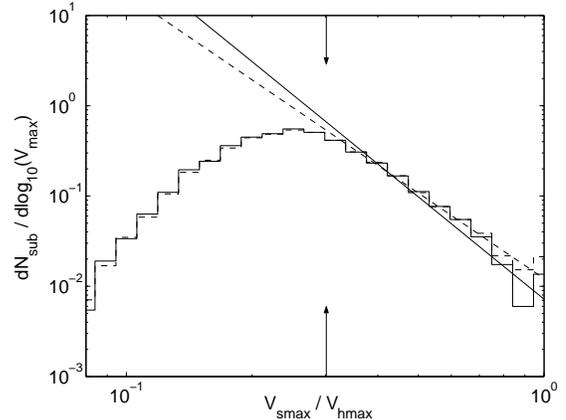}
\caption{Maximum circular velocity distribution of all the subhalos in
our coherent (dashed lines) and standard (solid) halo samples as a
function of the ratio of subhalo to halo maximum circular
velocity. Straight lines are the power law fits to the slope outwards
of the point indicated by the arrow, which is where we assume that the
distribution is not affected by the minimum subhalo mass to which we
can reliably measure subhalo maximum circular velocity (100 particles,
or $2.54\times 10^{11}  M_{\rm \sun}$. Slopes of the fit are $-3.13 \pm 0.11$
and $-3.66 \pm 0.30$ for the coherent and standard distributions
respectively.}
\label{fig:vcmax_func}
\end{figure}

Fig. \ref{fig:vcmax_func} shows the circular velocity function (CVF)
for subhalo maximum circular velocity as a fraction of halo maximum
circular velocity, $V_{smax} / V_{hmax}$. The maximum circular
velocity for both subhalos and halos was calculated using the 3
parameter fitting formula proposed by \cite{Stoehr:06} (hereafter,
STWS):
\begin{equation}
\log(\frac{V(x)}{V_{max}}) = -a[\log(x)]^2
\end{equation}
where $x = r/r_{vmax}$ and $V_{max}$, $r_{vmax}$ and $a$ are the
maximum circular velocity, the radius at which this is located, and the
width of the profile, respectively. We have adopted this profile because
the flexibility provided by the additional free parameter compared to
the NFW density profile \citep{Navarro:96}, results in a more accurate
measure of the maximum circular velocity of a halo. We find that the
STWS formula provides a good fit to the circular velocity profile of
subhalos of at least 100 particles. Hence, in
Fig. \ref{fig:vcmax_func} we only use subhalos of at least this mass.

We fit a power-law to the distribution for $V_{smax} > 0.3 V_{hmax}$,
where it is less affected by the incompleteness effects of a minimum
subhalo mass, obtaining a slope of $-3.13 \pm 0.11$ for the coherent
halo sample (dashed lines) and $-3.66 \pm 0.30$ for the standard
sample (solid lines). \citet{Reed:05} measure a slope of -4 for their
sample of 16 high resolution subhalos, with significant scatter. The
virial theorem gives the simple scaling relation, $M \propto
V_{max}^{\beta}$, where $\beta=3$ for isolated halos. However, for
subhalos having undergone mass loss, and because of a weak correlation
between halo mass and concentration, $\beta$ tends to vary between
3-4 \citep{Avila-Reese:99, Bullock:01b, Hayashi:03, Kravtsov:04,
Shaw:06}. Hence, the relative values we obtain for the slopes of the
SMF and CVF largely agree with what may be expected given the
relationship between $M_{vir}$ and $V_{max}$.

Overall, we again find a general good agreement between the two
samples, with evidence of a slightly higher number of high $V_{smax}$
subhalos in the coherent sample, resulting in the shallower
slope. There is also a tiny fraction of subhalos with a greater
maximum circular velocity than their host. These are very high mass
subhalos -- $M_{sub} \approx 0.38 M_{vir}$ -- of almost equal mass to
the mother halo itself. It is generally known that the maximum
circular velocity is a quantity that is fairly robust to changes in
subhalo mass due to dynamical friction and tidal stripping
(e.g. \cite{Kravtsov:04}). Hence, it is not necessarily surprising that
we should find high mass subhalos with $V_{smax}$ slightly greater
than their hosts.

\begin{figure}
\plotone{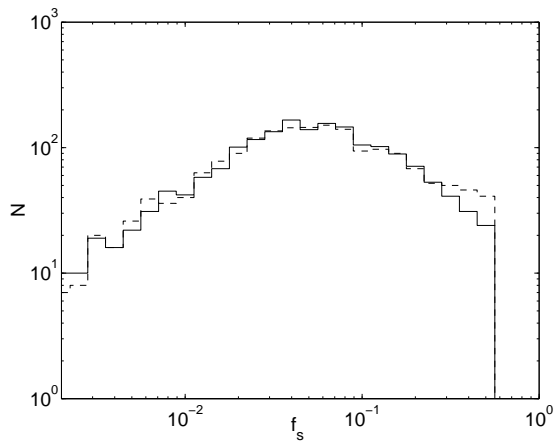}
\caption{Distribution of the fraction of mass contained within
substructure ($f_s$) for all the halos in our coherent (dashed lines)
and standard (solid) samples.}
\label{fig:frac_hist}
\end{figure}

In Fig. \ref{fig:frac_hist} we compare the distributions of the total
fraction of mass contained in substructure ($f_s$) for both our halo
samples. Aside from the slightly increased number of very high mass
substructures (relative to their host halo mass) discussed earlier,
there is a very good agreement between the two definitions of
substructure. The mean value of $f_s$ for the fully self-consistent
definition of substructure halo sample is $ 0.089 \pm 0.073$, compared
to $0.082 \pm 0.066$ for the standard definition sample (errors
correspond to one standard deviation). These values also agree
extremely well with those found in other studies \citep{DeLucia:04,
Gao:04, Gill:04b} who tend to find average values of $f_s$ in the
range 8-10\%.
\begin{figure}
\plotone{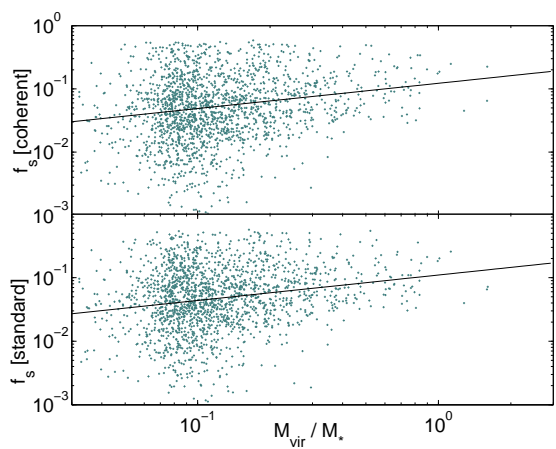}
\caption{Fraction of mass contained within substructure as a function
of halo mass, for our coherent halo sample (upper panel) and our
standard halo sample (lower). Solid lines are the best fit power law
to each distribution, where we obtain slopes of $0.44 \pm 0.09$ 
and $0.40 \pm 0.09$ for the
coherent and standard samples respectively. We take $M_*$ as
$8\times 10^{14}  M_{\rm \sun}$.}
\label{fig:mass_frac}
\end{figure}

In Fig. \ref{fig:mass_frac} we plot $f_s$ against halo mass for the
coherent (top panel) and the standard (lower panel) halo samples. It
is clear that for both samples that higher mass halos, on average,
contain a higher fraction of their mass in substructure.  To each
distribution we fit a power law, obtaining similar slopes of $0.44 \pm
0.09$ and $0.40 \pm 0.09$ for the coherent and standard samples,
respectively. This is an important result as it demonstrates that
cluster-mass halos are not self-similar. \cite{Gao:04} found a similar
result analyzing a reasonably large sample of halos over a wide mass
range.  \cite{vandenBosch:05} constructed a semi-analytical model to
calculate the subhalo mass function, adjusting the free parameters to
match the simulations of \citet{Gao:04}, \citet{Tormen:04} and
\citet{DeLucia:04}; they found the slope and normalization of the mass
function to be dependent on the ratio of halo mass to the
characteristic non-linear mass scale. In our previous paper
\citep{Shaw:06} we demonstrated that halos with a higher fraction of
their mass contained in substructure typically also have a lower
concentration, less spherical morphology and greater spin. We find the
same results apply to the halos in our coherent sample.

Overall, we find that accounting for all the forces on a substructure,
both internal (including the contribution of the `background'
particles to the binding energy) and external (incorporating the
tidally disruptive effect of the cluster mass), does
not produce a significantly different subhalo population from the
definition of substructure used in previous studies. In the following
section, we analyze and discuss the radial distributions of
substructures in our halo samples.

\subsection{Radial Distribution of Substructures}

The simplest assumption one may make is that the radial distribution
of satellite galaxies in a cluster follows the distribution of the
dark matter. However, a number of recent studies investigating the
radial distribution of subhalos in dissipationless simulations of
cluster mass halos have found that this distribution is significantly
less concentrated, and shallower in the inner regions, than that of
the dark matter \citep{Ghigna:98, Colin:99, Ghigna:00, Springel:01a,
DeLucia:04, Gao:04, Nagai:04, Reed:05,
Faltenbacher:06}. \citet{Diemand:04} performed convergence tests to
show that the discrepancy between the dark matter density and subhalo
radial distributions is likely not due to numerical resolution issues
in the inner regions of halos; rather, it is that the destructive
forces to which a subhalo is subjected (dynamical friction, tidal
shocks, merging) proceed more rapidly in the inner regions than at the
periphery of the system. \citet{Nagai:04} found in their simulations
that whilst the radial distribution of subhalos was significantly less
concentrated than that of the dark matter when selected using their
present-day gravitationally bound mass, their radial distribution
followed that of the dark matter much more closely when selected using
their mass at the time of accretion. By performing simulations
including gas dynamics, star formation and cooling, they also
demonstrate that the radial distributions of subhalos selected by
their stellar mass resulted in profiles similar in shape to that of
the dark matter. As stars are located in the core of subhalos, and are
therefore tightly bound, the stellar mass in a subhalo is a quantity
that should not vary substantially over time.

\begin{figure}
\plotone{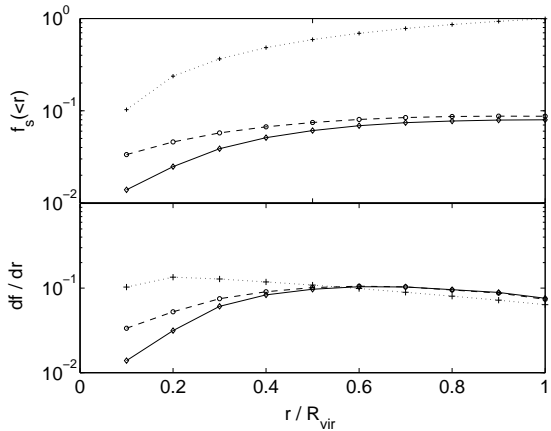}
\caption{({\it upper panel}) Cumulative fraction of mass in
substructure within spheres of radius $r/R_{vir}$, for our coherent
(dashed line) and standard (solid line) halo samples. Dotted line
denotes the total mass contained with radius $r$ as a fraction of the
total halo mass at the virial radius. ({\it lower}) The fraction of
mass in substructure in radial bins, $df/dr = (M_{sub}(r+dr) -
M_{sub}(r))/(M_{tot}(r+dr) - M_{tot}(r))$, for both halo samples. For
reference we also plot $(M_{tot}(r+dr) - M_{tot}(r))/M_{vir}$ (dotted
line).}
\label{fig:rad_frac}
\end{figure}

\begin{figure}
\plotone{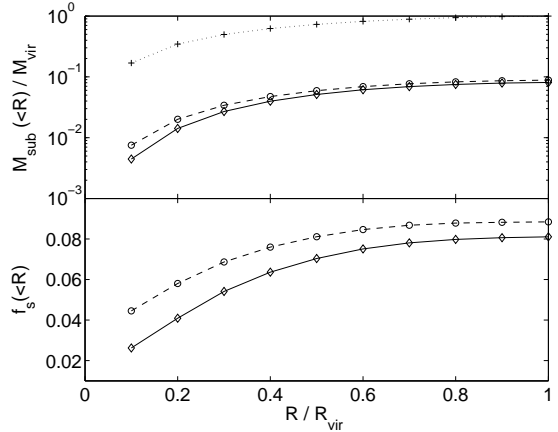}
\caption{({\it upper panel}) Cumulative projected subhalo mass for the
coherent (dashed lines) and standard (solid lines) halo samples as a
fraction of halo virial mass, $M_{sub}(<R)/M(R_{vir})$. Dotted line
denotes the total halo mass contained within successively increasing
fractions of the virial radius as a fraction of halo virial mass,
$M_{tot}(<R)/M(R_{vir})$. ({\it lower panel}) Cumulative projected
mass in substructure in circles of radius $R/R_{vir}$ as a fraction of
the total halo mass within the circular region,
$M_{sub}(<r)/M_{tot}(<r)$.}
\label{fig:rad_frac_proj}
\end{figure}

In order to compare the radial distributions of substructure in each
of our samples, in the upper panel of Fig. \ref{fig:rad_frac} we plot
the spherically averaged fraction of mass in substructure within a
given radius $r$, $f_s(<r/R_{vir})$.  This is calculated by dividing
the mass in substructures by the total mass inside spheres of
successively increasing fractions of the virial radius; we then take
the mean value of all the halos in each sample at each radii. The
circles/dashed line show the coherent halo sample, and the
diamonds/solid line show the standard sample. For reference, we also
plot the mean fraction of the total halo mass, 
$M(<r/R_{vir})/M(R_{vir})$
within each radius (crosses/dotted line). In
Fig. \ref{fig:rad_frac_proj} we plot the {\it projected} radial
distribution of the mass contained in substructure; this is calculated
by taking the mean of three orthogonal projections of each halo. In
the upper panel we plot the substructure mass within a projected
fraction of the halo virial radius, $R/R_{vir}$, as a fraction of halo virial
mass, or $M_{sub}(<R) / M(R_{vir})$, for the coherent (dashed line)
and standard (solid) halo samples. For reference, we also plot the
total halo mass within $R/R_{vir}$ as a fraction of $M_{vir}$ (dotted
line). In the lower panel we plot the substructure mass within
$R/R_{vir}$ as a fraction of the total halo mass within $R/R_{vir}$,
or $M_{sub}(<R) / M_{DM}(<R)$.

As found in previous studies, it is clear that the radial distribution
of substructure does not follow that of the dark matter 
in the inner regions of a cluster. We re-emphasize that
the minimum subhalo mass in each sample is 30 particles or $7.62\times
10^{10} h^{-1}  M_{\rm \sun}$ and the minimum cluster virial mass is 10,000
particles or $2.54\times 10^{13} h^{-1}  M_{\rm \sun}$. Outward of
$0.6R_{vir}$, the cumulative fraction of mass in substructure for both
samples remains constant, indicating that in the outer regions the
subhalos trace the overall dark matter distribution fairly
well. Within this radius, the mass in subhalos relative to the dark
matter decreases significantly. This is much more pronounced for the
standard halo sample, which drops off much more steeply than the
coherent sample. This is also shown in the lower panel of
Fig. \ref{fig:rad_frac}, where we plot the mean value of $f_s$ for
each sample in radial bins. This is calculated by dividing the mass in
substructure between $r$ and $r+dr$ by the total halo mass within the
same spherical annulus. For reference we also plot $(M_{tot}(r+dr) -
M_{tot}(r))/M_{vir}$ at each radii (dotted line). Outwards of $0.5R_{vir}$ the
results for each sample are very similar, but inside this radius there
is significantly more substructure in the coherent halo sample. Thus
in the inner regions of a halo, the binding effect of including the
background (or unbound particles) located within a subhalo is on
average slightly stronger than than the tidal forces from the cluster
core and other surrounding substructures.

\subsection{Subhalos of subhalos}\label{sec:subsub}

When a halo is accreted onto a more massive host, it may bring with it
its own internal hierarchy of substructures. This is observed by
\cite{Gill:04b}, who follow the temporal evolution of
satellites in their simulations as they are accreted by their
host. Assuming this internal structure survives the initial impact of
the disruptive environment within a cluster, we might expect that
more massive halos will contain several generations of subhalos. Both
\citet{Zentner:05} and \citet{Taylor:04a} try to account for
subhalos-of-subhalos during their construction of semi-analytic models
for the subhalo populations of dark matter halos.

We have performed a preliminary investigation of the amount of mass
that is contained in subhalos that are not immediate daughters of the
mother halo. \citet{Weller:05} describe in detail
how the family-tree style hierarchy of substructures in each halo is
calculated, depending on whether each subhalo is bound to the next
most massive structure in the vicinity. Due to the resolution limits
of our simulation, we are only able to resolve substructures down to a
mass of approximately $10^{-3} - 10^{-4}$ of the cluster halo virial
mass. However, this is good enough to ensure that many of the cluster
halos in each of our samples contain at least two generations of
substructures. Indeed, several halos in our sample contain a small
number of {\it third} generation subhalos.

We have previously defined $f_s$ as the fraction of halo mass
contained in substructure. We now also define $f_{ss}$ as the fraction
of mass in each subhalo contained in second generation subhalos. Of
course, the minimum subhalo mass in our simulation, $7.62 \times
10^{10}h^{-1}  M_{\rm \sun}$ or 30 particles, inhibits our ability to identify
sub-subhalos in subhalos more than it does to identify subhalos in
halos. Therefore we only include in our sample subhalos with masses of
at least 1\% that of their host halo, or $M_{sub} \ge 0.01\, M_{vir}$,
when calculating $f_s$, and second generation subhalos with $M_{ss}
\ge 0.01 \, M_{sub}$ when calculating $f_{ss}$. Additionally, we only
analyze the sub-subhalo populations of subhalos that satisfy the first
criterion, i.e. $M_{sub} \ge 0.01\, M_{vir}$. All halos and subhalos
that contain no subhalos or sub-subhalos, respectively, are
discarded. We therefore now have a consistent means of comparing the
populations of different generations of subhalos.

\begin{figure}
\plotone{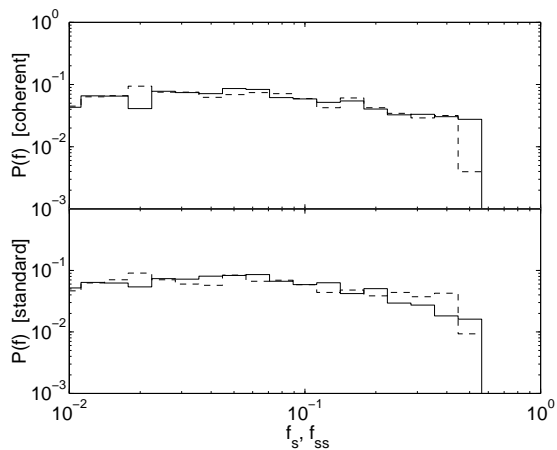}
\caption{Distribution of the fraction of the halo mass contained
within substructure, $f_s$ (solid lines) and the fraction of subhalo
mass contained within second generation subhalos, $f_{ss}$ (dashed lines) for our
coherent ({\it upper panel}) and standard ({\it lower panel}) halo
samples. Distributions are normalized by the total number of halos for
which $f_s$ -- and subhalos for which $f_{ss}$ -- is greater than
zero.}
\label{fig:subsub}
\end{figure}

In the upper panel of Fig. \ref{fig:subsub} we compare the
distribution of $f_s$ for all the halos in our coherent sample (solid
line) to the distribution of $f_{ss}$ for all the {\it subhalos} in
the entire sample (dashed line) in logarithmic bins. In the lower panel, we
plot the equivalent distributions for the standard halo sample. We
normalize each distribution so that it is expressed as a fraction of
the total number of halos with $f_s>0$ for the former, and the number
of halos with $f_{ss}>0$ for the latter. Hence, the distributions
represent the probability of a halo containing $f_s$ of its mass in
substructure, and the probability of a subhalo containing $f_{ss}$ of
its mass in sub-substructure, assuming that both $f_s$ and $f_{ss}$ are
greater than zero.

The similarity between the distributions of $f_{s}$ and $f_{ss}$ --
especially for the coherent sample -- is striking. It suggests a
self-similarity within the hierarchical structures of cluster mass
halos, with the 2nd generation of subhalos distributed in the 1st
generation, as the latter is in the cluster as a whole. This result is
both new and extremely interesting, especially considering that it
has been demonstrated that halos are not self-similar -- younger,
higher mass halos tend to contain more substructure than their older,
lower mass counterparts \citep{Gao:04, Diemand:04, Shaw:06}. However,
it may not entirely be surprising. Just as recently formed halos
contain a higher fraction of their mass in substructure, recently {\it
accreted} subhalos might be expected to be the same -- they have only
recently become exposed to the disruptive environment within a halo, and so
still retain a significant number of their own subhalos. Older
halos have had more time to disrupt the substructures they host, so
older subhalos will have lost a higher fraction of their own sub-subhalos
due to the same processes. We hope to investigate this further with higher
resolution simulations, enabling the identification of several
generations of subhalos, in the near future.

\section{DISCUSSION AND CONCLUSION}

In this paper we have presented a improved definition of subhalos in
dissipationless dark matter N-body simulations, based on the coherent
identification of their dynamically bound constituents. Whereas
previous methods determine the energetically bound components of a
subhalo while ignoring the contribution of particles in the halo that are
not geometrically or dynamically associated with it, our method
accounts for all the forces, both internal and external, exerted on a
subhalo. This is realized in two stages: first, when calculating the
energetically bound components of a subhalo, we include the
contribution to the gravitational potential of those particles that
were geometrically assigned to it, but that have been identified as
unbound in previous iterations of the calculation. In previous
studies, the contribution of unbound particles has been ignored. The
effect of including their forces is to increase the binding energy, and
therefore the mass of a subhalo. Second, we allow for the external
forces exerted on each subhalo by the rest of the particles in the
halo. This is achieved by approximating the tidal forces on each face
of a cube surrounding each subhalo. We then calculate whether the
tidal forces on each particle are sufficient to detach it from the
subhalo within a characteristic time, discarding those for which this
is the case. The effect of this stage therefore is to reduce the mass
of subhalos -- especially those near to the halo core or other dense
conglomerates of particles.

To demonstrate that our new {\it coherent} method of identifying
subhalos results in a more accurate measure of their bound mass than
the standard procedure, we tracked two samples of approximately 50
subhalos over six consecutive timesteps (from z = 0.13 to 0, or 1.17
Gyrs), identified using the coherent and standard methods
respectively. By checking whether particles identified as bound to a
subhalo (by either criterion) remained so, and whether nearby
particles that were found not to be bound did indeed move away, we
were able to assess how accurately each method identifies the bound
mass of a subhalo. We demonstrated that the standard method is less
successful at removing particles that are not bound to a subhalo. The
average fraction of a subhalos mass that was identified as bound by
the standard method but was found to have left the subhalo by the next
timestep was, on average, a factor of 1.29 greater than that measured
for the coherent method (9\% compared to 7\%). As expected, the
accuracy of both methods decreases when we compare subhalos to their
descendants several timesteps ahead; after 5 timesteps (1.17 Gyrs or
$\approx 14$ characteristic dynamical times) 28\% and 25\% of the
particles identified as `bound' by the standard and coherent methods
respectively, were found to have left the subhalo.

We then applied both the standard method of identifying subhalos and our
new coherent definition to a sample of 1838 virialized halos extracted
from a high resolution cosmological simulation. We found the subhalo
mass and maximum circular velocity distributions (relative to the mass
and maximum circular velocity of their hosts) for each sample to be
very similar. The slope of the high mass end of the subhalo mass
distribution was $-0.79 \pm 0.04$ and $-0.91 \pm 0.03$ for the
coherent and standard halo sample respectively. The slightly flatter
slope for the coherent sample was due to a small increase in the
number of very high mass subhalos, relative to their host. We obtained
equivalent slopes of $-3.13 \pm 0.11$ and $-3.66 \pm 0.30$ for the
subhalo maximum circular velocity distributions. The mean fraction of
mass contained in substructure, $f_s$, was $0.089 \pm 0.074$ for the
coherent halo sample and $0.082 \pm 0.066$ for the standard
sample. These results are all within the range of values found by
previous authors.

In agreement with \citet{Gao:04}, we find that $f_s$ tends to increase
with halo mass for both halo samples.  High mass halos ($M_{vir} >
M_*$) have formed more recently and have therefore have had less time
to disrupt the subhalos the halos they accreted. In our previous study
\citep{Shaw:06}, we demonstrated that -- for our {\it standard} halo
sample -- halos with a high $f_s$ are typically also less spherical,
have a lower concentration and a higher angular momentum than halos
with a lower fraction of their mass in substructure. We have verified
that the same is true for our coherent halo sample. These results
break the assumption that is adopted in some semi-analytic models
\citep[e.g.][]{Oguri:04} that dark matter halos are self-similar, with
galaxy mass halos resembling `miniature' cluster mass halos. Recently,
using a prescription for modelling the evolution of subhalos under the
influence of dynamical fraction and tidal stripping,
\citet{Zentner:05} and \citet{Taylor:04a} have developed semi-analytic
models that correctly predict the deviations from a self-similar
scaling of subhalo abundance with halo mass.

We then proceeded to investigate the radial distributions of
substructure in each sample. We found that within $60\%$ of the halo
virial radius the distribution of subhalos is significantly less
concentrated than that of the dark matter particles. However, this
effect was less pronounced for subhalos in the coherent halo sample,
indicating that in the central regions, the binding effect of
including the `background particles' in the potential calculation is
slightly stronger than the disruptive effect of the tidal forces. This
increases the mass of the subhalos in the inner regions of a halo
relative to their counterparts in the standard halo sample.

Overall, however, we found that the inclusion of the contribution of
`background' particles to the binding energy of a subhalo was offset
by the disruptive effect of the tidal forces from
particles outside a subhalo. As these two effects, on average, tend to
negate each other, we found that the subhalo populations were not
substantially different to those found using the standard definition of
substructure.

Finally, we performed a preliminary comparison of the subhalo
population of halos to the second generation, or {\it sub-subhalo},
populations of subhalos. As part of our subhalo identification scheme,
we construct for each halo a `family tree' of substructure, checking
to see whether smaller subhalos are bound to more massive
subhalos. Our results indicate that the distribution of 2nd generation
subhalos within their first generation hosts, mimics that of the
distribution of subhalos in the overall halo. Hence, we find there to
be a self-similar scaling between the populations of different
generations of subhalos within halos. This is a new and interesting
result, especially given the {\it lack} of self-similarity in the
overall subhalo populations of host halos of different mass. We intend
to investigate this further in the near future, with higher resolution
simulations, and thus with halos containing multiple generations of
subhalos.

\section{ACKNOWLEDGMENTS}
LDS acknowledges a PPARC student fellowship. This work was supported
by a grant of supercomputing time (grant number MCA04N002P) from the
National Center for Supercomputing Applications, and also used
computational facilities supported by NSF grant AST-0216105. This work
also made use of the COSMOS (SGI Altix 3700) supercomputer at DAMTP in
Cambridge and on the Sun Sparc-based Throughput Engine at the
Institute of Astronomy in Cambridge. Cosmos is a UK-CCC facility which
is supported by HEFCE and PPARC. We thank Scott Tremaine for helpful
discussions and the referee for many useful suggestions.


\input{ms.bbl}
\label{lastpage}
\end{document}

%% file: tab1.tex
\begin{table*}
\begin{center}
\begin{tabular}{| l | c c c | c c c |}
\hline
 & & & & & & \\
 & \multicolumn{3}{c|}{Coherent Analysis} & \multicolumn{3}{c|}{Standard Analysis} \\
 & & & & & & \\
\hline
 & & & & & & \\
Step  & $N_p$ & $f_s$ & Nsubs & $N_p$ & $f_s$ & Nsubs\\
 & & & & & & \\
\hline
 & & & & & & \\
Denmax &  472659 &  34.8\% &  2185 &  472659 &  34.8\% &  2185 \\
 & & & & & & \\
Subhalo unbinding and &  472659  &  8.4\% & 61 &  472659  & 4.75\% & 56 \\
Hyperstructures & & & & & & \\
 & & & & & & \\
Tidal Step & 472659 & 5.2\% & 58 & & n/a &  \\
 & & & & & & \\
Cluster unbinding and &  402411 & 4.5\% & 43 & 402534 & 4.0 \% & 41 \\
virial cut & & & & & & \\
\hline
\end{tabular}
\caption{Comparison between subhalo definitions of the substructure
content of Halo 1 after the completion of each step in the respective procedures.}
\label{tab:steps}
\end{center} 
\end{table*}

%% file: tab2.tex
\begin{table*}
\begin{center}
\begin{tabular}{| c c c c c c |}
\hline
 & & & & & \\
 step & z & $t_{lookback}$ & $M_{halo}$ & $N_{sub}$ (coh) &  $N_{sub}$ (std)\\
 & & [Gyrs] & [$h^{-1} M_{sun}$] & & \\
 & & & & & \\
\hline
 & & & & & \\
35 & 0.13 & 1.17 & $1.40 \times 10^{13}$ & 59 & 58 \\
 & & & & & \\
36 & 0.1 & 0.92 & $1.41 \times 10^{13}$ & 56 & 56 \\
 & & & & & \\
37 & 0.07 & 0.66 & $1.43 \times 10^{13}$ & 56 & 55 \\
 & & & & & \\
38 & 0.05 & 0.48 & $1.44 \times 10^{13}$ & 57 & 54 \\
 & & & & & \\
39 & 0.02 & 0.19 & $1.45 \times 10^{13}$ & 60 & 50 \\
 & & & & & \\
40 & 0 & 0.00 & $1.46 \times 10^{13}$ & 60 & 52 \\
 & & & & & \\
\hline
\end{tabular}
\caption{Mass and the number of subhalos (as identified by the
standard (std) and coherent (coh) methods) in a halo followed over the
final 6 steps of a $256^3$ particle simulation (discussed in Section
\ref{sec:trackhalos}).}
\label{tab:tsteps}
\end{center}
\end{table*}

%% file: ms.bbl
\begin{thebibliography}{59}
\expandafter\ifx\csname natexlab\endcsname\relax\def\natexlab#1{#1}\fi

\bibitem[{{Amara} {et~al.}(2006){Amara}, {Metcalf}, {Cox}, \&
  {Ostriker}}]{Amara:06}
{Amara}, A., {Metcalf}, R.~B., {Cox}, T.~J., \& {Ostriker}, J.~P. 2006, \mnras,
  367, 1367

\bibitem[{{Avila-Reese} {et~al.}(1999){Avila-Reese}, {Firmani}, {Klypin}, \&
  {Kravtsov}}]{Avila-Reese:99}
{Avila-Reese}, V., {Firmani}, C., {Klypin}, A., \& {Kravtsov}, A.~V. 1999,
  \mnras, 310, 527

\bibitem[{{Bertschinger} \& {Gelb}(1991)}]{Bertschinger:91}
{Bertschinger}, E. \& {Gelb}, J.~M. 1991, Computers in Physics, 5, 164

\bibitem[{Binney \& Tremaine(1987)}]{Binney:87a}
Binney, J. \& Tremaine, S. 1987, Galactic Dynamics (Princeton University Press,
  Princeton, NJ, USA)

\bibitem[{{Bode} {et~al.}(2001){Bode}, {Ostriker}, \& {Turok}}]{Bode:01}
{Bode}, P., {Ostriker}, J., \& {Turok}, N. 2001, ApJ, 556, 93

\bibitem[{Bode \& Ostriker(2003)}]{Bode:03a}
Bode, P. \& Ostriker, J.~P. 2003, ApJS, 145, 1B

\bibitem[{{Bullock} {et~al.}(2001){Bullock}, {Kolatt}, {Sigad}, {Somerville},
  {Kravtsov}, {Klypin}, {Primack}, \& {Dekel}}]{Bullock:01b}
{Bullock}, J.~S., {Kolatt}, T.~S., {Sigad}, Y., {Somerville}, R.~S.,
  {Kravtsov}, A.~V., {Klypin}, A.~A., {Primack}, J.~R., \& {Dekel}, A. 2001,
  \mnras, 321, 559

\bibitem[{{Col{\'{\i}}n} {et~al.}(1999){Col{\'{\i}}n}, {Klypin}, {Kravtsov}, \&
  {Khokhlov}}]{Colin:99}
{Col{\'{\i}}n}, P., {Klypin}, A.~A., {Kravtsov}, A.~V., \& {Khokhlov}, A.~M.
  1999, \apj, 523, 32

\bibitem[{{Das} \& {Ostriker}(2006)}]{Das:06}
{Das}, S. \& {Ostriker}, J.~P. 2006, \apj, 645, 1

\bibitem[{{Davis} {et~al.}(1985){Davis}, {Efstathiou}, {Frenk}, \&
  {White}}]{Davis:85}
{Davis}, M., {Efstathiou}, G., {Frenk}, C.~S., \& {White}, S.~D.~M. 1985, \apj,
  292, 371

\bibitem[{{De Lucia} {et~al.}(2004){De Lucia}, {Kauffmann}, {Springel},
  {White}, {Lanzoni}, {Stoehr}, {Tormen}, \& {Yoshida}}]{DeLucia:04}
{De Lucia}, G., {Kauffmann}, G., {Springel}, V., {White}, S.~D.~M., {Lanzoni},
  B., {Stoehr}, F., {Tormen}, G., \& {Yoshida}, N. 2004, \mnras, 348, 333

\bibitem[{{Diemand} {et~al.}(2004){Diemand}, {Moore}, \& {Stadel}}]{Diemand:04}
{Diemand}, J., {Moore}, B., \& {Stadel}, J. 2004, \mnras, 352, 535

\bibitem[{{Eisenstein} \& {Hut}(1998)}]{Eisenstein:98}
{Eisenstein}, D.~J. \& {Hut}, P. 1998, \apj, 498, 137

\bibitem[{{Faltenbacher} \& {Diemand}(2006)}]{Faltenbacher:06}
{Faltenbacher}, A. \& {Diemand}, J. 2006, \mnras, 369, 1698

\bibitem[{{Gao} {et~al.}(2004){Gao}, {White}, {Jenkins}, {Stoehr}, \&
  {Springel}}]{Gao:04}
{Gao}, L., {White}, S.~D.~M., {Jenkins}, A., {Stoehr}, F., \& {Springel}, V.
  2004, \mnras, 355, 819

\bibitem[{{Gelb} \& {Bertschinger}(1994)}]{Gelb:94}
{Gelb}, J.~M. \& {Bertschinger}, E. 1994, \apj, 436, 467

\bibitem[{{Ghigna} {et~al.}(1998){Ghigna}, {Moore}, {Governato}, {Lake},
  {Quinn}, \& {Stadel}}]{Ghigna:98}
{Ghigna}, S., {Moore}, B., {Governato}, F., {Lake}, G., {Quinn}, T., \&
  {Stadel}, J. 1998, \mnras, 300, 146

\bibitem[{{Ghigna} {et~al.}(2000){Ghigna}, {Moore}, {Governato}, {Lake},
  {Quinn}, \& {Stadel}}]{Ghigna:00}
---. 2000, \apj, 544, 616

\bibitem[{{Gill} {et~al.}(2004{\natexlab{a}}){Gill}, {Knebe}, \&
  {Gibson}}]{Gill:04a}
{Gill}, S.~P.~D., {Knebe}, A., \& {Gibson}, B.~K. 2004{\natexlab{a}}, \mnras,
  351, 399

\bibitem[{{Gill} {et~al.}(2004{\natexlab{b}}){Gill}, {Knebe}, {Gibson}, \&
  {Dopita}}]{Gill:04b}
{Gill}, S.~P.~D., {Knebe}, A., {Gibson}, B.~K., \& {Dopita}, M.~A.
  2004{\natexlab{b}}, \mnras, 351, 410

\bibitem[{{Hagan} {et~al.}(2005){Hagan}, {Ma}, \& {Kravtsov}}]{Hagan:05}
{Hagan}, B., {Ma}, C.-P., \& {Kravtsov}, A.~V. 2005, \apj, 633, 537

\bibitem[{{Hayashi} {et~al.}(2003){Hayashi}, {Navarro}, {Taylor}, {Stadel}, \&
  {Quinn}}]{Hayashi:03}
{Hayashi}, E., {Navarro}, J.~F., {Taylor}, J.~E., {Stadel}, J., \& {Quinn}, T.
  2003, \apj, 584, 541

\bibitem[{{Helmi} {et~al.}(2002){Helmi}, {White}, \& {Springel}}]{Helmi:02}
{Helmi}, A., {White}, S.~D., \& {Springel}, V. 2002, \prd, 66, 063502

\bibitem[{{Hennawi} {et~al.}(2005){Hennawi}, {Dalal}, {Bode}, \&
  {Ostriker}}]{Hennawi:05}
{Hennawi}, J.~F., {Dalal}, N., {Bode}, P., \& {Ostriker}, J.~P. 2005,
  {astro-ph/0506171}

\bibitem[{{Huchra} \& {Geller}(1982)}]{Huchra:82}
{Huchra}, J.~P. \& {Geller}, M.~J. 1982, \apj, 257, 423

\bibitem[{{Kang} {et~al.}(2005){Kang}, {Jing}, {Mo}, \& {B{\"o}rner}}]{Kang:05}
{Kang}, X., {Jing}, Y.~P., {Mo}, H.~J., \& {B{\"o}rner}, G. 2005, \apj, 631, 21

\bibitem[{{Kazantzidis} {et~al.}(2004){Kazantzidis}, {Mayer}, {Mastropietro},
  {Diemand}, {Stadel}, \& {Moore}}]{Kazantzidis:04}
{Kazantzidis}, S., {Mayer}, L., {Mastropietro}, C., {Diemand}, J., {Stadel},
  J., \& {Moore}, B. 2004, \apj, 608, 663

\bibitem[{{Kim} \& {Park}(2006)}]{Kim:06}
{Kim}, J. \& {Park}, C. 2006, \apj, 639, 600

\bibitem[{{Klypin} {et~al.}(1999){Klypin}, {Gottl{\" o}ber}, {Kravtsov}, \&
  {Khokhlov}}]{Klypin:99}
{Klypin}, A., {Gottl{\" o}ber}, S., {Kravtsov}, A.~V., \& {Khokhlov}, A.~M.
  1999, \apj, 516, 530

\bibitem[{{Kravtsov} {et~al.}(2004{\natexlab{a}}){Kravtsov}, {Berlind},
  {Wechsler}, {Klypin}, {Gottl{\"o}ber}, {Allgood}, \&
  {Primack}}]{Kravtsov:04b}
{Kravtsov}, A.~V., {Berlind}, A.~A., {Wechsler}, R.~H., {Klypin}, A.~A.,
  {Gottl{\"o}ber}, S., {Allgood}, B., \& {Primack}, J.~R. 2004{\natexlab{a}},
  \apj, 609, 35

\bibitem[{{Kravtsov} {et~al.}(2004{\natexlab{b}}){Kravtsov}, {Gnedin}, \&
  {Klypin}}]{Kravtsov:04}
{Kravtsov}, A.~V., {Gnedin}, O.~Y., \& {Klypin}, A.~A. 2004{\natexlab{b}},
  \apj, 609, 482

\bibitem[{{Lacey} \& {Cole}(1994)}]{Lacey:94}
{Lacey}, C. \& {Cole}, S. 1994, \mnras, 271, 676

\bibitem[{{Lahav} {et~al.}(1991){Lahav}, {Lilje}, {Primack}, \&
  {Rees}}]{Lahav:91}
{Lahav}, O., {Lilje}, P.~B., {Primack}, J.~R., \& {Rees}, M.~J. 1991, \mnras,
  251, 128

\bibitem[{Mao {et~al.}(2004)Mao, Jing, Ostriker, \& Weller}]{Mao:04}
Mao, S., Jing, Y.-P., Ostriker, J.~P., \& Weller, J. 2004, Astrophys. J., 604,
  L5

\bibitem[{{Moore} {et~al.}(1999){Moore}, {Ghigna}, {Governato}, {Lake},
  {Quinn}, {Stadel}, \& {Tozzi}}]{Moore:99a}
{Moore}, B., {Ghigna}, S., {Governato}, F., {Lake}, G., {Quinn}, T., {Stadel},
  J., \& {Tozzi}, P. 1999, ApJL, 524, L19

\bibitem[{{Moore} {et~al.}(1996){Moore}, {Katz}, \& {Lake}}]{Moore:96}
{Moore}, B., {Katz}, N., \& {Lake}, G. 1996, \apj, 457, 455

\bibitem[{{Nagai} \& {Kravtsov}(2005)}]{Nagai:04}
{Nagai}, D. \& {Kravtsov}, A.~V. 2005, \apj, 618, 557

\bibitem[{{Navarro} {et~al.}(1996){Navarro}, {Frenk}, \& {White}}]{Navarro:96}
{Navarro}, J.~F., {Frenk}, C.~S., \& {White}, S.~D.~M. 1996, \apj, 462, 563

\bibitem[{{Neyrinck} {et~al.}(2005){Neyrinck}, {Gnedin}, \&
  {Hamilton}}]{Neyrinck:04}
{Neyrinck}, M.~C., {Gnedin}, N.~Y., \& {Hamilton}, A.~J.~S. 2005, \mnras, 356,
  1222

\bibitem[{{Oguri} \& {Lee}(2004)}]{Oguri:04}
{Oguri}, M. \& {Lee}, J. 2004, \mnras, 355, 120

\bibitem[{{Okamoto} \& {Habe}(1999)}]{Okamoto:99}
{Okamoto}, T. \& {Habe}, A. 1999, \apj, 516, 591

\bibitem[{{Reed} {et~al.}(2005){Reed}, {Governato}, {Quinn}, {Gardner},
  {Stadel}, \& {Lake}}]{Reed:05}
{Reed}, D., {Governato}, F., {Quinn}, T., {Gardner}, J., {Stadel}, J., \&
  {Lake}, G. 2005, \mnras, 359, 1537

\bibitem[{{Shaw} {et~al.}(2006){Shaw}, {Weller}, {Ostriker}, \&
  {Bode}}]{Shaw:06}
{Shaw}, L.~D., {Weller}, J., {Ostriker}, J.~P., \& {Bode}, P. 2006, \apj, 646,
  815

\bibitem[{{Springel} {et~al.}(2001){Springel}, {White}, {Tormen}, \&
  {Kauffmann}}]{Springel:01a}
{Springel}, V., {White}, S.~D.~M., {Tormen}, G., \& {Kauffmann}, G. 2001,
  \mnras, 328, 726

\bibitem[{Stadel {et~al.}(1997)Stadel, Katz, Weinberg, \&
  Hernquist}]{Stadel:97a}
Stadel, J., Katz, N., Weinberg, D.~H., \& Hernquist, L. 1997, {\tt
  www-hpcc.astro.washington.edu/tools/skid.html}

\bibitem[{{Stadel}(2001)}]{Stadel:01}
{Stadel}, J.~G. 2001, Ph.D.~Thesis

\bibitem[{{Stoehr}(2006)}]{Stoehr:06}
{Stoehr}, F. 2006, \mnras, 365, 147

\bibitem[{{Summers} {et~al.}(1995){Summers}, {Davis}, \& {Evrard}}]{Summers:95}
{Summers}, F.~J., {Davis}, M., \& {Evrard}, A.~E. 1995, \apj, 454, 1

\bibitem[{{Taffoni} {et~al.}(2003){Taffoni}, {Mayer}, {Colpi}, \&
  {Governato}}]{Taffoni:03}
{Taffoni}, G., {Mayer}, L., {Colpi}, M., \& {Governato}, F. 2003, \mnras, 341,
  434

\bibitem[{{Tasitsiomi} {et~al.}(2004){Tasitsiomi}, {Kravtsov}, {Wechsler}, \&
  {Primack}}]{Tasitsiomi:04b}
{Tasitsiomi}, A., {Kravtsov}, A.~V., {Wechsler}, R.~H., \& {Primack}, J.~R.
  2004, \apj, 614, 533

\bibitem[{{Taylor} \& {Babul}(2004)}]{Taylor:04a}
{Taylor}, J.~E. \& {Babul}, A. 2004, \mnras, 348, 811

\bibitem[{{Tormen} {et~al.}(2004){Tormen}, {Moscardini}, \&
  {Yoshida}}]{Tormen:04}
{Tormen}, G., {Moscardini}, L., \& {Yoshida}, N. 2004, \mnras, 350, 1397

\bibitem[{{Vale} \& {Ostriker}(2006)}]{Vale:06}
{Vale}, A. \& {Ostriker}, J.~P. 2006, \mnras, 371, 1173

\bibitem[{{van den Bosch} {et~al.}(2005){van den Bosch}, {Tormen}, \&
  {Giocoli}}]{vandenBosch:05}
{van den Bosch}, F.~C., {Tormen}, G., \& {Giocoli}, C. 2005, \mnras, 359, 1029

\bibitem[{{van Kampen}(1995)}]{vanKampen:95}
{van Kampen}, E. 1995, \mnras, 273, 295

\bibitem[{{Wambsganss} {et~al.}(2004){Wambsganss}, {Bode}, \&
  {Ostriker}}]{Wambsganss:04}
{Wambsganss}, J., {Bode}, P., \& {Ostriker}, J.~P. 2004, Ap.\ J.\ Lett., 606,
  L93

\bibitem[{{Weller} {et~al.}(2005){Weller}, {Ostriker}, {Bode}, \&
  {Shaw}}]{Weller:05}
{Weller}, J., {Ostriker}, J.~P., {Bode}, P., \& {Shaw}, L. 2005, \mnras, 364,
  823

\bibitem[{{White}(1976)}]{White:76}
{White}, S.~D.~M. 1976, \mnras, 177, 717

\bibitem[{{Zentner} {et~al.}(2005){Zentner}, {Berlind}, {Bullock}, {Kravtsov},
  \& {Wechsler}}]{Zentner:05}
{Zentner}, A.~R., {Berlind}, A.~A., {Bullock}, J.~S., {Kravtsov}, A.~V., \&
  {Wechsler}, R.~H. 2005, \apj, 624, 505

\end{thebibliography}
